\documentclass[apj]{emulateapj}
\usepackage{natbib}



\lefthead{SONGAILA \& COWIE}
\righthead{Variation of the Fine-Structure Constant}

\begin{document}

\title{Constraining the Variation of the Fine Structure Constant with Observations of Narrow Quasar Absorption Lines\altaffilmark{1}}

\author{A. Songaila \& L. L. Cowie}

\affil{Institute for Astronomy, University of Hawaii, 2680 Woodlawn Drive, Honolulu, HI 96822}

\altaffiltext{1}{Based in part on data obtained at the W. M. Keck
Observatory, which is jointly operated by
the California Institute of Technology, the University of
California, and the National Aeronautics and Space Administration.}

\vskip 1in

\slugcomment{Accepted for publication in  {\it Astrophysical Journal\/} }

\begin{abstract}
The unequivocal demonstration of temporal or spatial variability in a fundamental
constant of nature would be of enormous significance.  Recent
attempts to measure the variability of the fine structure constant $\alpha$ 
over cosmological time, using
high resolution spectra of high redshift quasars observed with 10m class telescopes, have produced conflicting results.  
We use the Many Multiplet (MM) method with \ion{Mg}{2} and \ion{Fe}{2} lines on very high signal-to-noise, high resolution ($R = 72,000$) Keck HIRES spectra of eight narrow quasar absorption systems. 
We consider both systematic uncertainties in spectrograph wavelength calibration and also velocity offsets introduced by complex velocity structure in even apparently simple and weak narrow lines and analyze their effect on claimed variations in $\alpha$. 
We find no significant change in $\alpha$, 
$\Delta\alpha/\alpha =(0.43\pm0.34) \times 10^{-5}$, in the redshift range
$z=0.7-1.5$, where this includes both statistical and systematic errors.  

We also show that the scatter in measurements of $\Delta\alpha/\alpha$\ arising from absorption line structure can be considerably larger than assigned statistical errors even for apparently simple and narrow absorption systems.  We find a null result of $\Delta\alpha/\alpha = (-0.59\pm0.55) \times 10^{-5}$\ in a system at $z=1.7382$ using lines of \ion{Cr}{2}, \ion{Zn}{2} and \ion{Mn}{2}, whereas using \ion{Cr}{2} and \ion{Zn}{2} lines in a system at $z=1.6614$\ 
we find a systematic velocity trend which, if interpreted as a shift in $\alpha$, would correspond to $\Delta\alpha/\alpha = (1.88\pm0.47) \times 10^{-5}$, where both results include both statistical and systematic errors.  This latter result is almost certainly caused by varying ionic abundances in subcomponents of the line: using 
\ion{Mn}{2}, \ion{Ni}{2} and \ion{Cr}{2} in the analysis changes the result to
$\Delta\alpha/\alpha = (-0.47\pm 0.53) \times 10^{-5}$.  Combining the \ion{Mg}{2} and \ion{Fe}{2} results with estimates based on \ion{Mn}{2}, \ion{Ni}{2} and \ion{Cr}{2} gives $\Delta\alpha/\alpha = (-0.01\pm 0.26) \times 10^{-5}$. 
We conclude that spectroscopic measurements  of quasar absorption lines are not yet capable of unambiguously detecting  variation in $\alpha$ using the MM method.

\end{abstract}

\section{Introduction}
\label{secintro}

There has been much interest over the last decade in the possibility that
fundamental dimensionless constants may evolve or in some way be a
product of the history of the region in which they lie \citep{uzan03}.
In the case of the fine-structure constant, $\alpha$, one group has
claimed to see such time variation \citep{murphy03a,murphy03b,murphy04,webb03}
and spatial variation \citep{webb11,webb14,king12} 
using the ``many multiplet'' (MM) method of analyzing quasar absorption
line spectra, which they pioneered \citep{webb99,dzuba99a,dzuba99b}.   However, this positive result is
limited to one group and more stringent null results, using
the same methodology, have been reported \citep{srianand04,chand04,chand05,chand06,levshakov06,molaro13}. 
The early results have been criticized by \citet{murphy08,murphy08b} who claim that their quoted significance is too high though not, in the case of the Chand et al.\ result, consistent with evidence for varying $\alpha$ \citep{srianand07,murphy08}.  
Investigations of the change in $\alpha$\ by comparing redshifted radio lines also do not show evidence of time variation \citep[e.g.][]{cowie95,carilli00,kanekar10,kanekar12,levshakov12}. These results provide an alternative source of evidence for a non-varying $\alpha$ that has different  systematic effects to the MM method, though the interpretation depends on assuming a low value for the change in $\mu \equiv m_e/m_p$ for which there is direct evidence at $ z < 0.9$ \citep{kanekar11,bagdonaite13a,bagdonaite13b}. (See e.g. \citet{rahmani13} for a summary of results at higher redshift).  
The discrepancy in
published results could be reconciled if systematic effects dominate
the measurements, leading to underestimates of the errors.  
In the present paper we consider both systematic uncertainties in spectrograph wavelength calibration and also velocity offsets introduced by complex velocity structure in even apparently simple and weak narrow lines and analyze their effect on claimed variations in $\alpha$.

The current observational situation regarding variation in $\alpha$\
is far from clear.  Terrestrial laboratory and geophysical
measurements show no sign of any time variation on both
long and short time scales.  Laboratory measurements \citep [e.g.][]{uzan03} of
the stability of atomic clocks are consistent with no instantaneous
time variation at the present time, while analysis of
the OKLO natural fission reactor 
places a limit of $(-0.11 \le \Delta\alpha/\alpha \le 0.24) \times 10^{-7}$  
over a time span of 2 Gyr \citep{gould06}. 
Astronomical measurements, on the other hand, have so far given
both null results and also the first reported detections of variation.  A
number of groups have looked for variation in $\alpha$\ over cosmic
time in quasar absorption lines at high redshift using various
methodologies.  All such measurements based on comparing the relative
wavelength separation of the two members of an alkaline doublet
observed at high redshift in a quasar absorption
system \citep{savedoff56} are consistent with no evolution 
\citep{potekhin94,cowie95,varshalovich96,murphy01a,chand05}.
It was
only with the advent of the more sensitive 
MM method \citep{webb99,dzuba99a,dzuba99b} 
that a significant change at the level of
$\Delta\alpha/\alpha = (-0.57\pm0.11) \times 10^{-5}$\ was
claimed over the redshift range $0.2 < z < 4.2$ \citep{murphy04}.

The enormous advantage of the MM method is its sensitivity: it is
sensitive to $\alpha^{-1}\Delta\alpha$\ at a level of about $10^{-5}$
in an individual quasar absorption line system, making it about an
order of magnitude better than the alkaline doublet method. On the
other hand, this sensitivity is bought at the price of much greater
required control of systematic effects, among
the most important of which are the accuracy of the wavelength
calibrations in the echelle spectrographs being used for the
measurements, and the assumptions of uniform velocity
structure in the various ion species being compared and uniform
spatial abundance patterns \citep{bahcall04,uzan03,lopez02,prochaska03,rodriguez06}.
The wavelength calibration is particularly critical since, in contrast to the alkaline doublet method, the lines being compared in the MM method can have wide wavelength separations.
While these systematic effects have been extensively investigated and modeled \citep{murphy03b,murphy04}, it is
still likely that they do in fact dominate the measurement error and
are responsible for the discrepancies in published results.  

\begin{figure}[h]
\figurenum{1}
\includegraphics[width=3.9in,angle=90,scale=0.75]{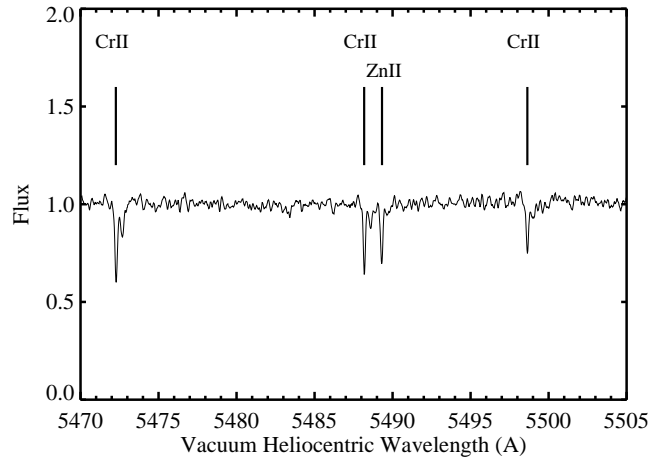}
\caption{Segment of the spectrum of HE1104-1805A
covering the \ion{Cr}{2} and \ion{Zn}{2} lines discussed in the text
showing how close in wavelength these lines are.
The spectrum is a 340-minute exposure obtained with the HIRES
spectrograph on the Keck I 10m telescope using a $0.57^{\prime\prime}$
slit. The full spectrum
covers the wavelength range from 4000 to $7800~{\rm\AA}$ with a
resolution of 72,000 measured from the sky lines. The wavelength
was calibrated with a 4th-order polynomial fit to a ThAr calibration
lamp taken before and after the exposure and absolutely calibrated
using night sky lines in the spectrum as discussed in the text.
The final wavelengths are given as vacuum heliocentric. The
signal-to-noise ratio is approximately 80 per resolution element in this
wavelength region.  
\label{fig:lines}
}
\end{figure}

A suitable choice of ionic species can minimize or exclude some of the
potential systematic errors associated with the MM method.  In
particular, concerns about systematics in the wavelength calibration
can be reduced by using lines of singly ionised Cr and Zn to measure
individual offsets since these lines are neighboring and interspersed
with one another (Figure~\ref{fig:lines}).  Since the effect of a
change in $\alpha$ is large and in the opposite sense for the two
species \citep{dzuba02}, 
these two ions also provide an optimally
sensitive measurement. For the neighboring \ion{Cr}{2} $2026~\rm\AA$\
and the \ion{Zn}{2} $2026~\rm\AA$\ lines, a change of 
$\Delta\alpha /\alpha = 10^{-5}$ produces a relative offset of $0.48~{\rm km\
s}^{-1}$\ \citep{dzuba02}.  
This is large enough that with a simple sharp absorption
line we can hope to measure the offset directly.  
A crucially important estimate of the magnitude of systematic errors in
the MM method comes from having the required sensitivity and spectral
resolution to fully analyze a {\it single} system at one
redshift and map the velocity shifts from ion to ion \citep{levshakov06}.  
This is what we do here for two such systems, at $z = 1.7382$ in
the quasar HS1946+7658 and at $z=1.6614$\ 
in the quasar
HE1104$-$1805A.

However, suitable Cr and Zn systems are rare and most of the MM measurements have been made using \ion{Fe}{2} and \ion{Mg}{2} lines, which have two disadvantages.
First, the offset between \ion{Fe}{2} $2600~\rm\AA$
and $2342~\rm\AA$ and the \ion{Mg}{2} $2800~\rm\AA$ doublet 
is smaller ($0.21~{\rm km\ s}^{-1}$ for $\Delta\alpha /\alpha = 10^{-5}$) 
than for the Cr and Zn lines;  and second, these systems have a much wider
wavelength separation. We shall show here that it is extremely
hard using the HIRES spectrograph on the Keck I telescope to obtain the necessary wavelength accuracy for this type
of measurement.

%
%

\section{Data and Data Reduction}
\label{secdata}

For the present study we chose three quasars from our 
existing data base of high resolution quasar spectra \citep{songaila05} 
which appeared optimal for measuring the fine structure variation.
These are
HE1104-1805A ($z_{em} = 2.31$), HS1700+6416 ($z_{em} = 2.73$), and
HS1946+7658 ($z_{em} = 3.02$).  These three quasars are very bright, 
allowing us to obtain high signal-to-noise observations
even at very high spectral resolution ($R = 72,000$) and 
their spectra all contain narrow, strong (but not saturated)
\ion{Mg}{2} lines with corresponding strong \ion{Fe}{2}. 
In addition HE1104-1805A and HS1946+7658 contain damped Lyman $\alpha$\ (DLA) 
absorbers with simple sharp and unsaturated metal lines that are ideal for measuring
the fine structure variation with multiple lines and, in particular,
ZnII and CrII. 
HE1104-1805 is a gravitationally lensed system with multiple
images \citep{wisotzki93}, 
of which the brightest (image ``A'') is
extremely luminous. The $z=1.6614$\ DLA
absorber has intermediate strength \ion{Cr}{2} and \ion{Zn}{2}
lines in its spectrum ($N({\rm CrII}) \sim 7 \times 10^{12}\ {\rm cm}^{-2}$ and $N({\rm ZnII}) \sim 2.4 \times 10^{12}\ {\rm cm}^{-2}$; Figure~\ref{fig:lines}).  Furthermore, the dominant
feature in the weaker singly-ionized absorption lines of the
$z=1.6614$ system is extremely sharp ($b \sim 4\ {\rm km\ s}^{-1}$) and was only marginally resolved in
our earlier moderately high resolution ($R = 35,000$)
observations. These earlier observations already suggested there was a
velocity offset between the \ion{Cr}{2} and \ion{Zn}{2} lines in this
system. HS1946+7658 also has an isolated narrow system 
corresponding to the DLA absorber at $z=1.7382$
which is seen in the \ion{Cr}{2}, \ion{Zn}{2} and \ion{Mn}{2} lines,
though this system is weaker than that in HE1104-1805A and some useful
lines lie in the Ly$\alpha$ forest.  

The upgrade of the CCDs in HIRES spectrograph on the Keck~I 10m telescope 
in 2004 provided finer pixel sampling
and allowed the instrument to achieve proper sampling 
of the high spectral resolution that could be obtained with narrow slits.
We therefore began an intensive observing program on 
these three quasars using a slit width of $0.574^{\prime\prime}$, corresponding to 
a spectral resolution of $R = 72,000$, 
in which we paid careful attention to the wavelength calibration.  The seeing was generally larger than the slit width so the slit illumination should be uniform.  
We will discuss this in detail in \S\ref{secwave}.
For each observation, 80-- or 100--minute 
exposures were obtained in two 40-- or 50--minute segments to allow 
for processing to remove cosmic rays.  ThAr calibration
lamps and in some case Iodine cell exposures of a
white dwarf standard were taken before and after each pair of 
quasar exposures (i.e. every 80 minutes).  
The observations are summarized in 
Table~\ref{tab_obs}.  Each quasar was observed 
in multiple grating settings to provide complete wavelength coverage.  
This also allowed us to test how reproducible wavelengths
were in different configurations of the instrument.  The
total exposure times are 340 minutes for HE1104-1805A, 280 minutes
for HS1700+6416, and 600 minutes for HS1946+7658.


\begin{deluxetable}{lcll}
\tablecaption{Log of Observations \label{tab_obs}}
\tablehead{\colhead{Quasar}  &
\colhead{\ \ \ $z_{em}$\ \ } &
\colhead{Date (UT)} &
\colhead{\quad Exp. (s)}
}
\startdata
HE 1104$-$1805A & 2.31  & 2005 Dec 27 & \quad $3 \times 2400$\\
                              &           & 2009 Jan 31  & \quad $1 \times 2400$\\
                              &          & 2005 Dec 27 & \quad $2 \times 2400$\\
                              &           & 2008 Jan 3   & \quad $2 \times 2400$\\
HS 1700+6416      & 2.73  & 2005 July 2  & \quad $4 \times 2400$\\
                              &          & 2005 July 3  & \quad $3 \times 2400$\\
HS 1946+7658      & 3.02  & 2004 Sep 10  & \quad $6 \times 2400$\\
                              &          & 2005 July 3  & \quad $2 \times 2400$\\
                              &          & 2006 Aug 29 & \quad $3 \times 2400$\\
                              &          & 2005 July 2  & \quad $2 \times 2400$\\
                              &           & 2006 Aug 30 & \quad $2 \times 2400$
\enddata
\end{deluxetable}

\begin{figure}[h]
\figurenum{2}
\includegraphics[width=3.9in,angle=90,scale=0.75]{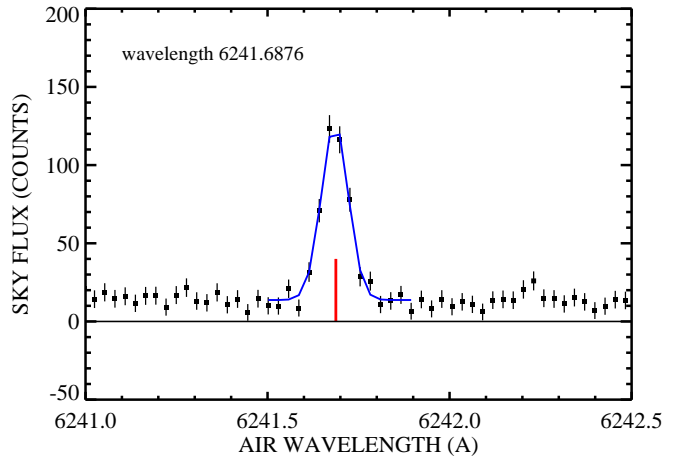}
\caption{Fit to a sky line at 6241.6876~\AA.  (This is actually a double but the lines are close enough to be indistinguishable).  The blue line is the fit to the sky, which is shown by the black squares.  The red line marks the wavelength of the sky line.  The wavelength scale is in the air system and corresponds to the average of the ThAr calibrations before and after the observation.
\label{fig:one_sky}
}
\end{figure}

The observations were obtained with a 2-point binning
in the spatial direction (corresponding to a binned
pixel size of $0.24^{\prime\prime}$) and no binning in the spectral
direction.
 Each spectrum was extracted by order using an IDL
procedure written for the purpose. The images were
initially flattened using normalized exposures of
a quartz flat lamp observed in the same spectrograph
configuration. In each order the
spatial distortion of the spectrum was measured from
white dwarf calibration stars taken in the same configuration;  these were also used to remove the
blaze function of the instrument. Sky subtraction was
performed by interpolating the sky measured on each side
of the spectrum and a vector corresponding to the
sky level at the position of the spectrum was stored for each order. Each spectrum was
then extracted using a 5 binned spatial pixel window centered
on the distorted spatial position at that wavelength to
obtain the total counts at that position. Cosmic rays
were rejected by comparing the spectra corresponding
to the sub-exposures and a final summation of the spectra
and the corresponding sky was then made and stored. The ThAr calibration
spectra were measured over the same positions used  to
extract the spectra and also stored by order. The raw counts
in the spectrum plus sky were used to compute the expected noise at
each pixel assuming a read noise of 6 electrons per pixel and a
gain setting of 2.4.  The final spectral resolution was measured from the sky lines and
is consistent with the nominal value of $R = 72,000$. We show a Gaussian fit to a sample sky line at 6241~\AA\  in Figure~\ref{fig:one_sky}.   The full width half maximum corresponds to 3.2 spectral pixels, giving a reasonable sampling.  In the following analysis we use the laboratory wavelengths of relevant ions summarized in \citet{murphy14}.

%
%

%
%

\begin{figure}[h]
\figurenum{3}
\includegraphics[width=3.9in,angle=0,scale=0.7]{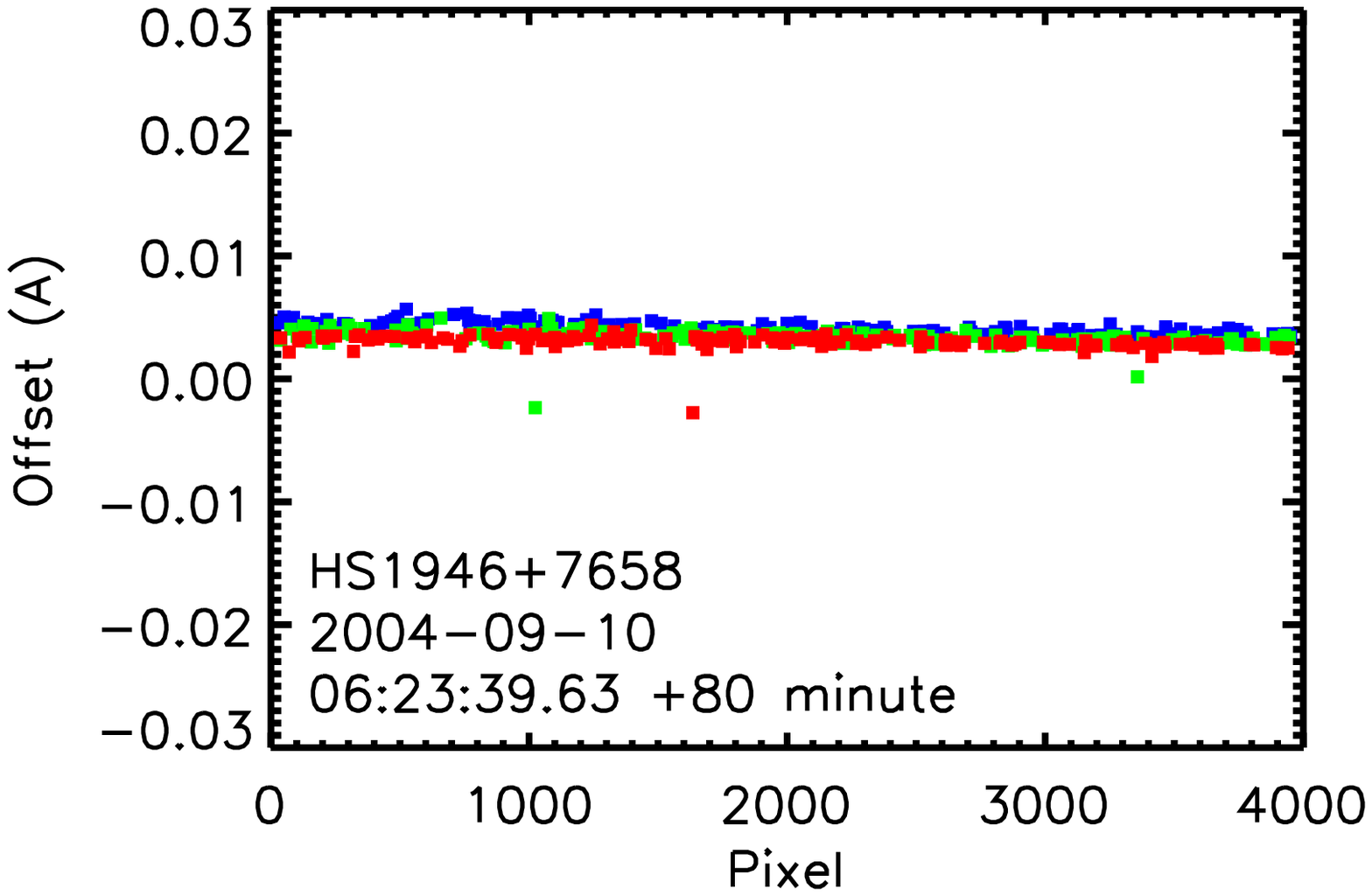}
\includegraphics[width=3.9in,angle=0,scale=0.7]{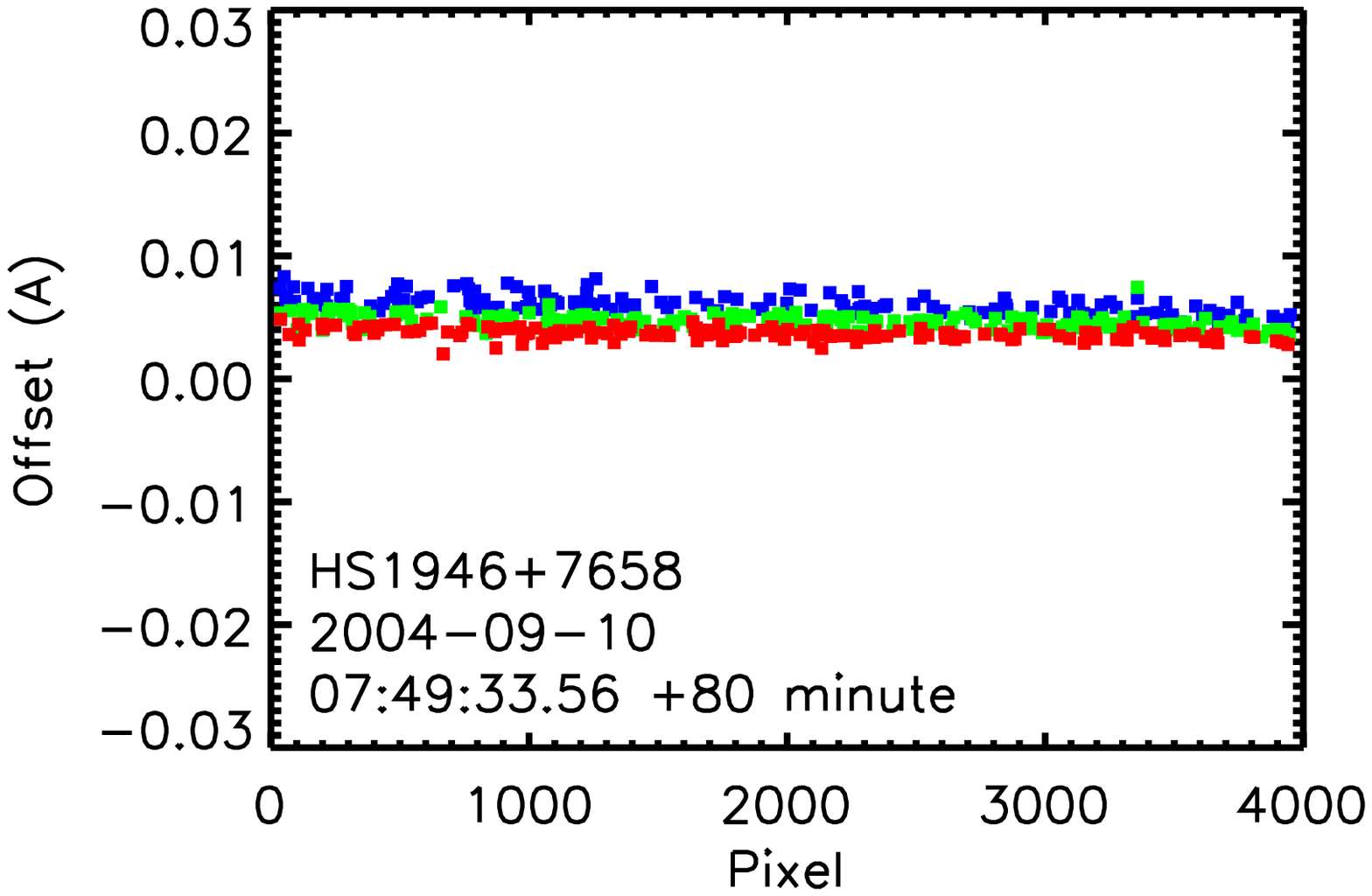}
\includegraphics[width=3.9in,angle=0,scale=0.7]{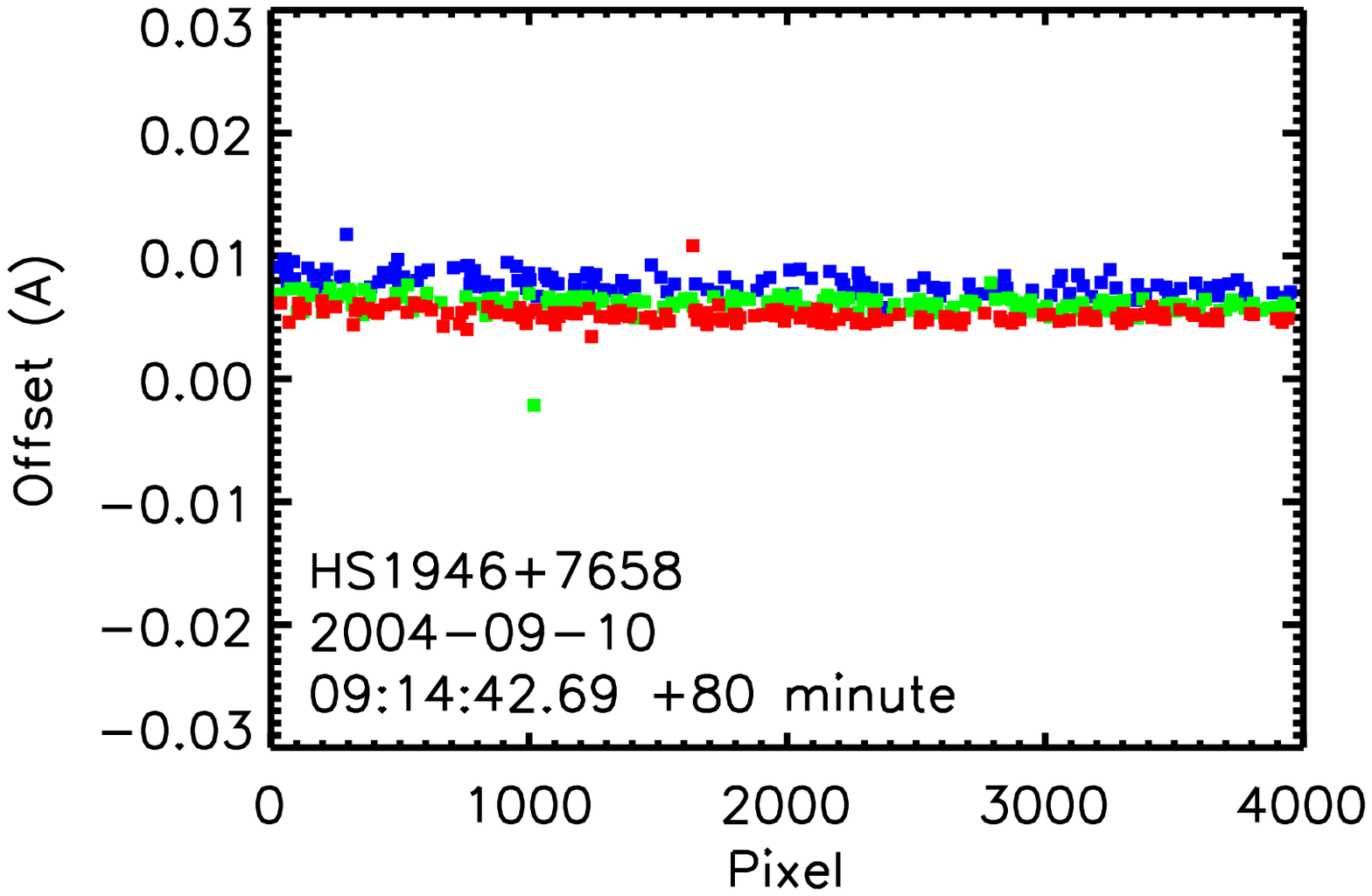}
\includegraphics[width=3.9in,angle=0,scale=0.7]{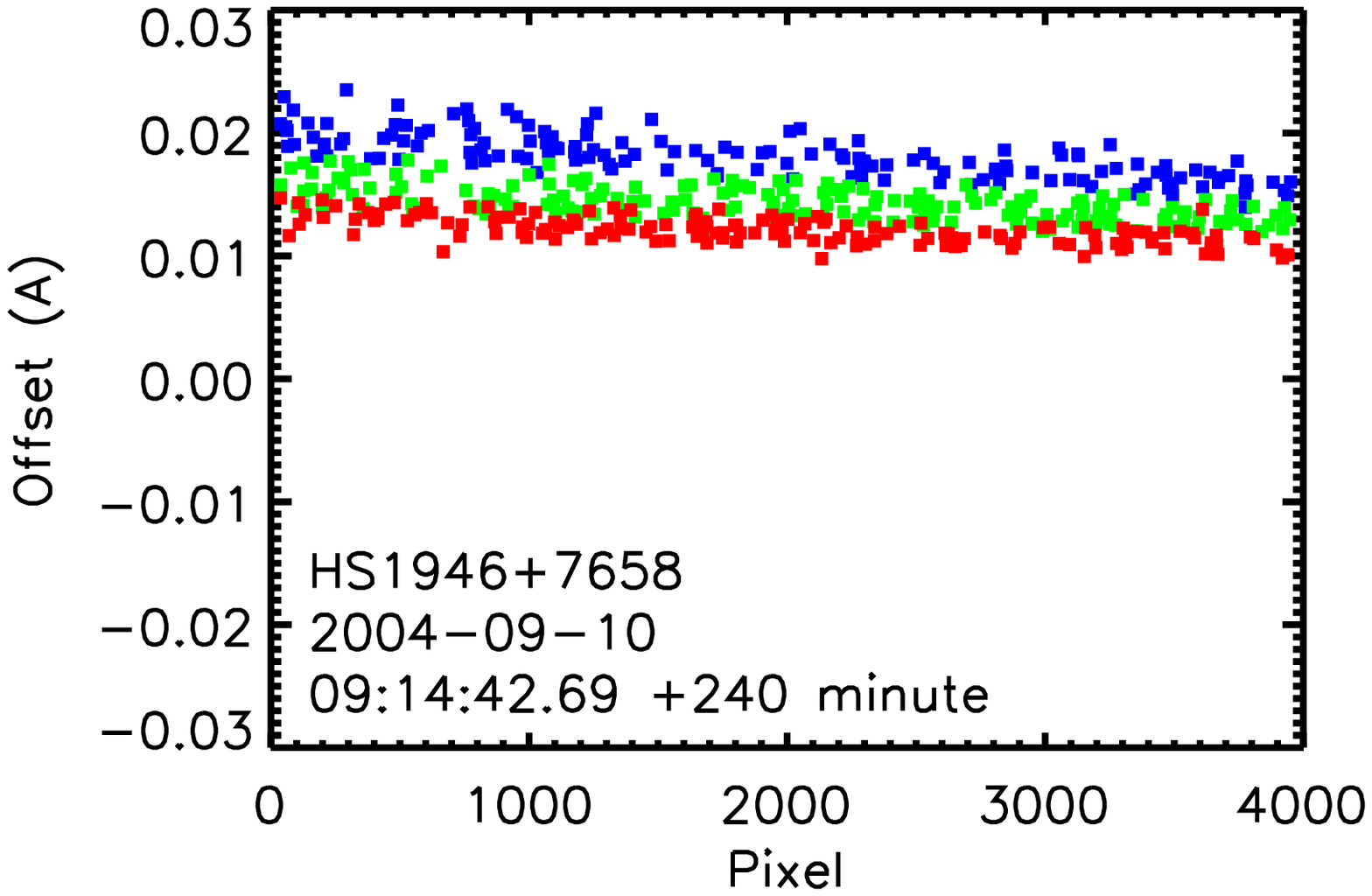}
\caption{Calibration drift as a function of time.  In each panel we show the difference in wavelength between the ThAr lines measured before and after the object exposure.  Each point shows the directly measured offset  of a single ThAr line with the orders color coded by wavelength from longest wavelengths (red) through intermediate wavelengths (green) to the shortest wavelengths (blue).  The drift in offset continues smoothly as a function of time (in the first three panels the text shows the time of the second ThAr exposure) resulting in a large offset for the full period of the observations (bottom panel).  The offset is differential with wavelength (see Figure~4) and also shows a gradient along each order.
\label{fig:drift}
}
\end{figure}

\section{Wavelength calibration}
\label{secwave}

The accuracy of the wavelength calibration is a crucial constraint in achieving the sensitivity necessary to measure evolution in the fundamental constants, particularly if widely separated lines such as \ion{Mg}{2} and \ion{Fe}{2} are being used, and requires the adoption of very careful observing techniques and attention to systematics.  This has previously been discussed for the HIRES spectrograph by \citet{griest10}, 
who used iodine cell exposures to evaluate the systematics of ThAr lamp calibration techniques, and \citet{thompson09}, 
who presented a calibration procedure for the VLT/UVES spectrograph that removes systematics previously noted in the  pipeline procedure by \citet{murphy07}.  
To obtain the wavelength calibration, we made ThAr lamp exposures before and after each object exposure, and also took a number of exposures each night of white dwarf standards with the iodine cell in place.  Since Griest et al.\ have made a thorough study of the iodine cell calibration, we will not discuss this issue any further here but concentrate on how well the calibration can be made using only ThAr lines.  
It should be noted that many of the archival observations used in previous analyses of the variation in the fine-structure constant did not aim for precision calibration.  The present discussion relates to observations specifically designed to obtain the most accurate calibrations possible with the ThAr lamp.
We first discuss the drift in wavelength calibration exhibited by the before and after ThAr lamp spectra, and then analyze the accuracy of the best possible ThAr calibrations using night sky lines that are present in the object spectra, and also using the multiple measurements of the various lines furnished by the independent observations, and comparing the velocities of doublets seen in the spectra.

%
%

\begin{figure}[h]
\figurenum{4}
\includegraphics[width=5in,angle=0,scale=0.7]{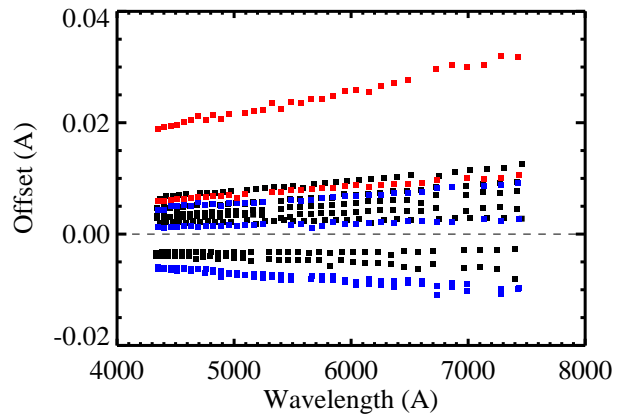}
\caption{The average wavelength displacement in each order as a function of the central wavelength of the order.  Black squares show data for HS 1946+7658, red for HE 1104$-$1805A and blue for HS 1700+6416.
\label{fig:drift_orders}
}
\end{figure}

\begin{figure*}[h]
\figurenum{5}
\includegraphics[width=3.8in,angle=90,scale=0.7]{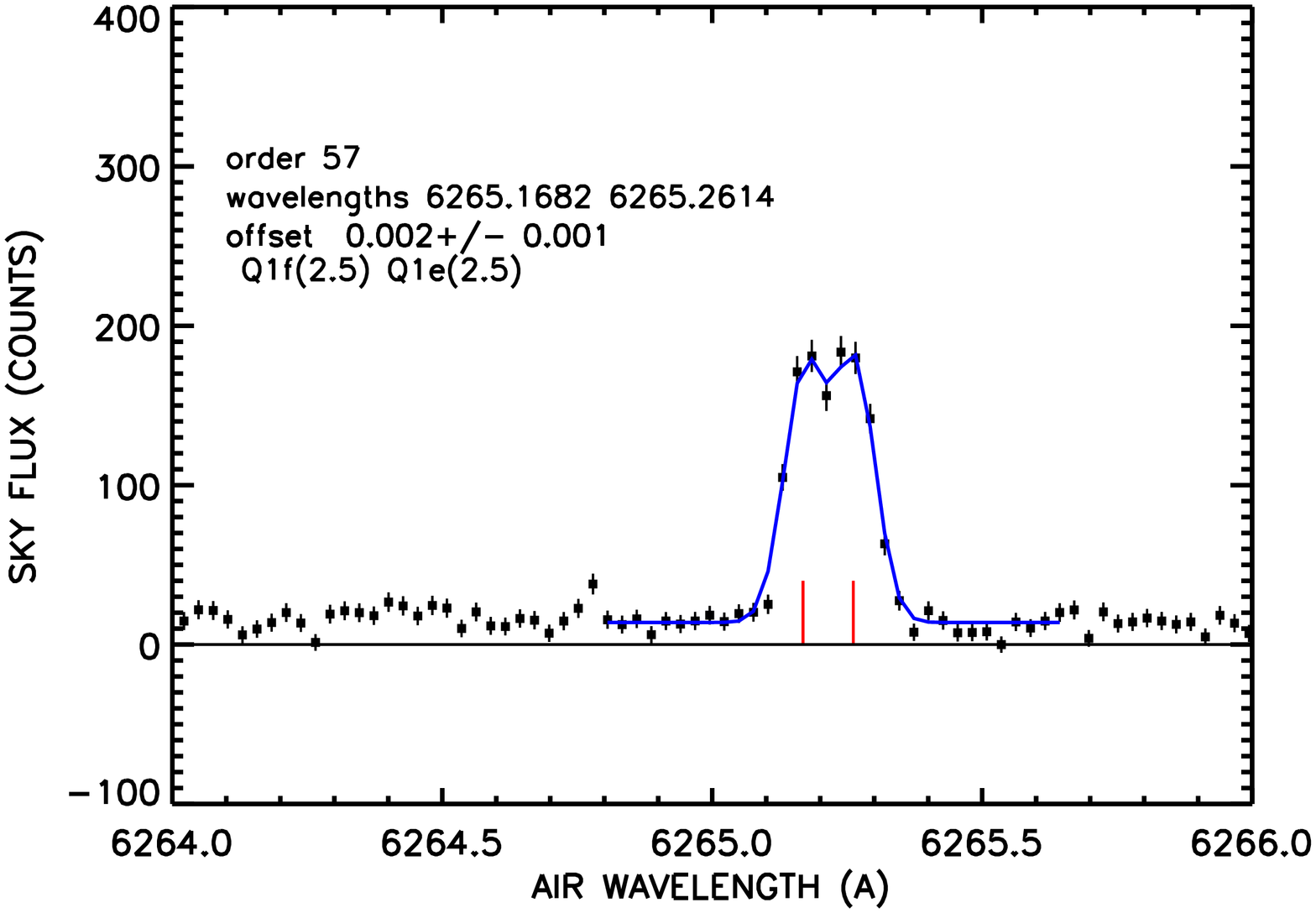}
\includegraphics[width=3.8in,angle=90,scale=0.7]{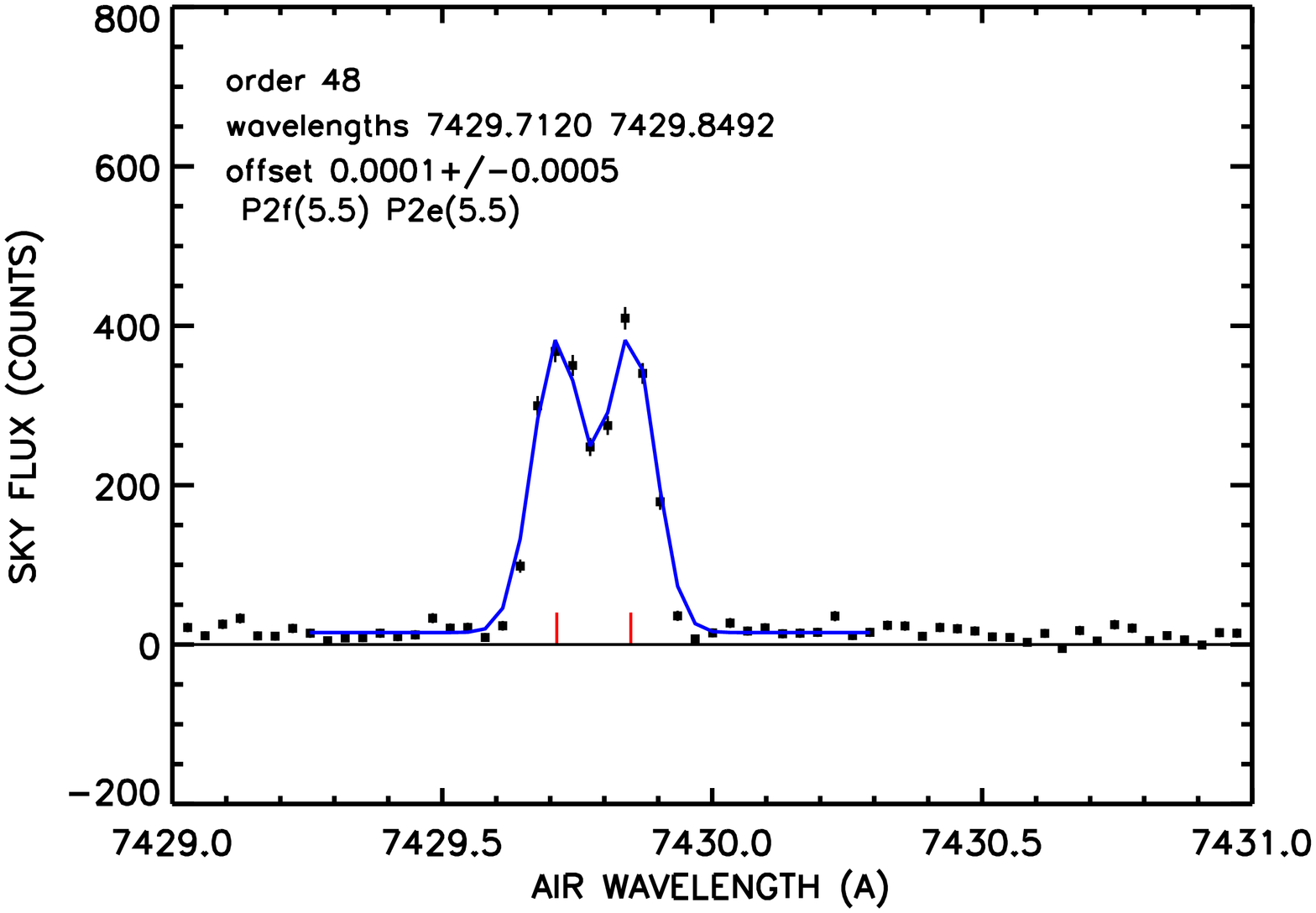}
\includegraphics[width=3.8in,angle=90,scale=0.7]{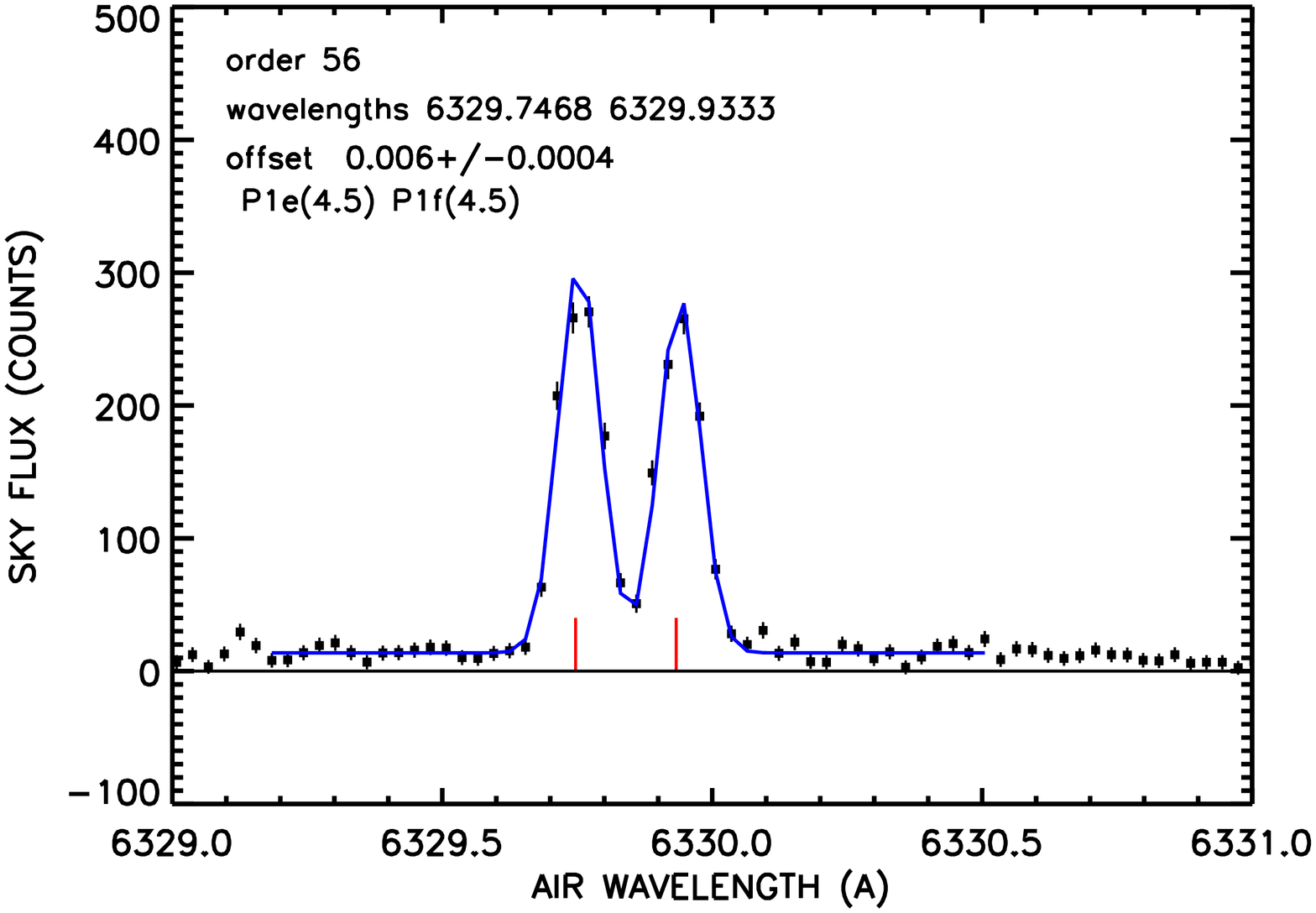}
\includegraphics[width=3.8in,angle=90,scale=0.7]{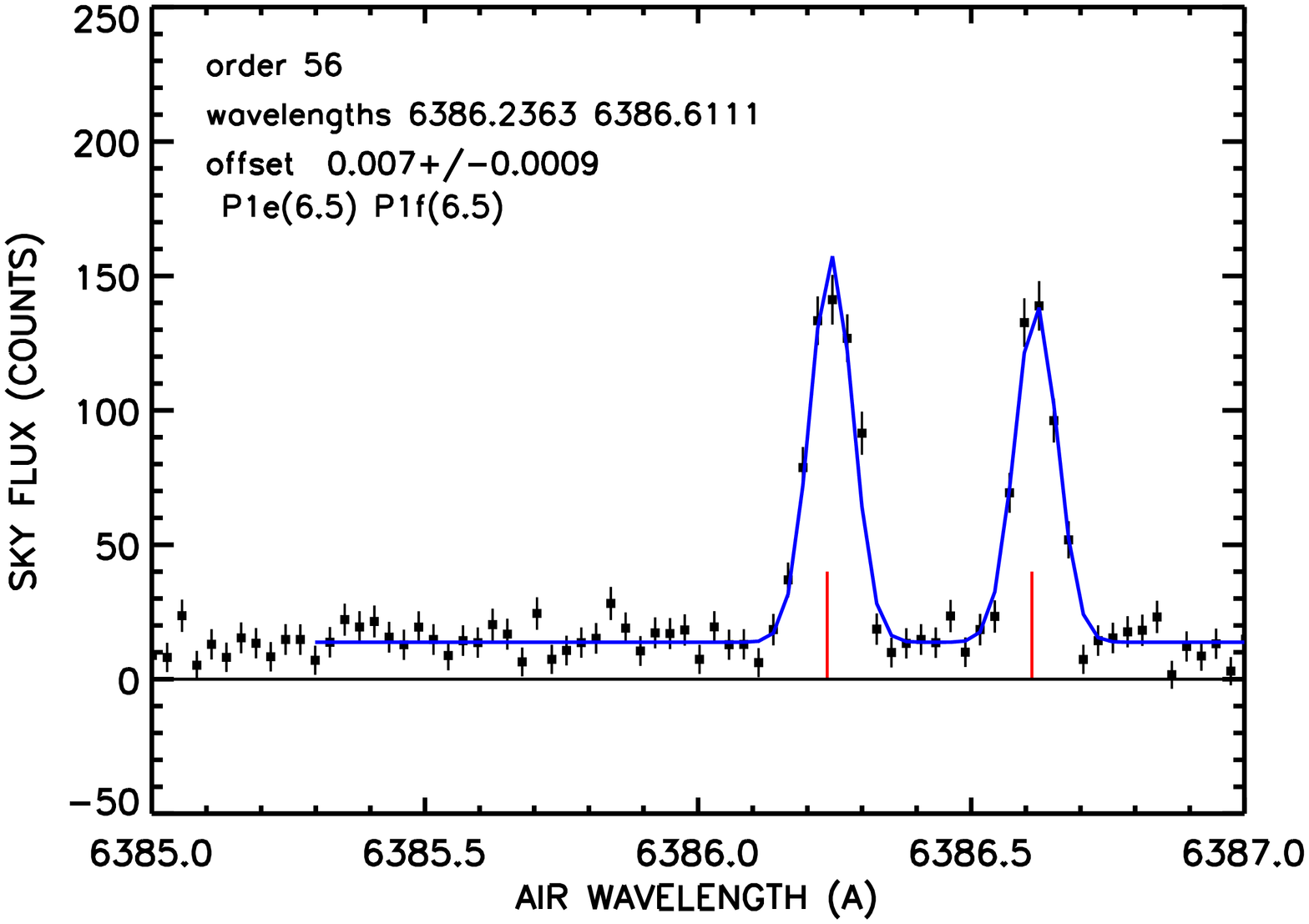}
\caption{Examples of night sky lines from the September 2004 
  observations of
HS1946+7658. The spectrum (black squares) is shown in the observed air
wavelength frame corresponding to the ThAr calibration
taken just before the observation. The wavelengths of
the two components of the sky line, split by $\Lambda$\ splitting, are shown with the red vertical marks.
The blue curve shows a Gaussian fit to the lines computed with the full width of the lines fixed to the instrument resolution of $R = 72,000$\  and the offset between the lines fixed to the theoretical separation.  The panels  show examples of splittings ranging from marginally resolved to a case where the splitting is 
four times the resolution.
The offsets of the sky lines from the true position are indicated in the text description in each panel.
\label{fig:skyline}
}
\end{figure*}

\subsection{ThAr lamp}
\label{tharcal}

A preliminary wavelength calibration was made using the
ThAr calibration lamp exposures made before and after each quasar exposure (i.e. with a typical 80 minute separation).  In each order we fitted Gaussians to between 15 and 20 lines
uniformly covering the wavelength range. We then used a 
fourth-order polynomial to obtain the wavelength fit.
The typical dispersion in the ThAr lines relative to
the wavelength solution is 0.0016\AA.

In order to measure the drift in the calibration we measured the shifts in individual ThAr lines between the before and after calibrations.  Examples of these measurements are shown in Figure~\ref{fig:drift}.  The most pronounced effect is an overall wavelength offset,  up to 0.025\AA\ at 6000\AA\ over an 80-minute observing period in an extreme case though the more typical value is $\sim 0.005$\AA .  
In Figure~\ref{fig:drift} we show that the effect is smoothly increasing as a function of time for a sequence of exposures, suggesting that it is caused by a regular change in the optical path, possibly driven by changes in the environment of the instrument. These overall shifts have no effect on the measurement of $\Delta \alpha/\alpha$, which depends on the relative offset of line pairs.  However, the shifts also result in differential offsets along the orders and in the cross-dispersed direction (Figure~\ref{fig:drift}).  The relative offset of the orders as a function of wavelength is shown in Figure~\ref{fig:drift_orders} for all of our ThAr pairs.  Both negative and positive absolute offsets are seen and the corresponding gradient is linearly related to the absolute offsets as 
$d({\delta \lambda} )/d{\lambda} \propto 1.5 \times 10^{-6}\, \delta \lambda\ (6000\rm\AA )$.  
Typical end-to-end wavelength shifts are $\sim 0.0015$\AA\ along the orders though in the most extreme case the value rises to $\sim 0.0050$\AA .

Under most conditions these offsets will scatter equally to negative and positive values and contribute only to the systematic errors.  However, depending on the cause of the drift and the observer procedure, they could result in systematic effects.  For example, if the offset were driven by temperature changes, which are more often negative through the night, and if observers normally take ThAr exposures prior to the on-sky observation, then there could be a systematic shift in $\Delta \alpha$.  However we assume here that the offsets simply add to the error budget. 

The actual errors will depend on the observing procedure but if we assume that the data are being calibrated with a ThAr exposure taken 80 minutes away from the observation (which may be fairly typical) then  we can use the current data to estimate the systematic error.  For a system at $z = 1$, the dispersion in the measured offset between \ion{Fe}{2} 2344\AA\ and \ion{Mg}{2} 2800\AA\ corresponds to an r.m.s. error of $4.5 \times 10^{-6}$\ in $\Delta \alpha/\alpha$\ for an individual measurement.

\subsection{Sky lines}
\label{skycal}

We can also use the night sky lines in the spectra of the quasars to test the accuracy of the final wavelength 
solution and to check if there are residual distortions in the
solution arising from the different optical path followed
by the ThAr calibration lamp versus sky objects illuminating the slit.  Since we used a very narrow slit (in all cases narrower than the typical seeing) the slit was approximately uniformly illuminated, and the sky lines should provide a good calibration.  For our ThAr wavelength scale we used the average of the wavelength solutions computed from the before and after ThAr wavelength fits, which should remove much of the drift discussed in the previous subsection.

We selected appropriate night sky lines from the list tabulated by \citet{osterbrock96} 
in the wavelength range $5500$ -- $7500~\rm\AA$.  To obtain sufficient accuracy, we recalculated the wavelengths of all lines from the term values in \citet{abrams94}.  
We fitted Gaussians to all sky lines using the line-fitting program MPFIT \citep{markwardt09}.  
Examples of the Gaussian fits are shown as the blue solid lines in Figure~\ref{fig:skyline}, plotted over the spectrum (black squares and error bars), shown in the observed air wavelength frame corresponding to the ThAr calibration taken immediately before the object exposure.  The OH sky lines are doublets owing to $\Lambda$-splitting.  The theoretical wavelengths of the $\Lambda$-split components are shown by the red lines.  We fitted Gaussians 
with the full width of the lines fixed to the measured instrument resolution of $R = 72,000$\  and the wavelength offset between the $\Lambda$\ - split components fixed to the theoretical separation. Figure~\ref{fig:skyline} shows  examples of sky lines with $\Lambda$-splitting ranging from just resolved to about four times the instrumental resolution.  The offsets of the ThAr-measured wavelengths of the sky lines from the true position range from $10^{-3}~\rm\AA$\ to $7 \times 10^{-3}~\rm\AA$.  These offsets are significantly larger than the statistical errors expected in the fits, which are typically in the range $5 \times 10^{-4}$\ -- $10^{-3}$\AA.

\begin{figure}[h]
\figurenum{6}
  \includegraphics[width=3in,angle=90,scale=0.9]{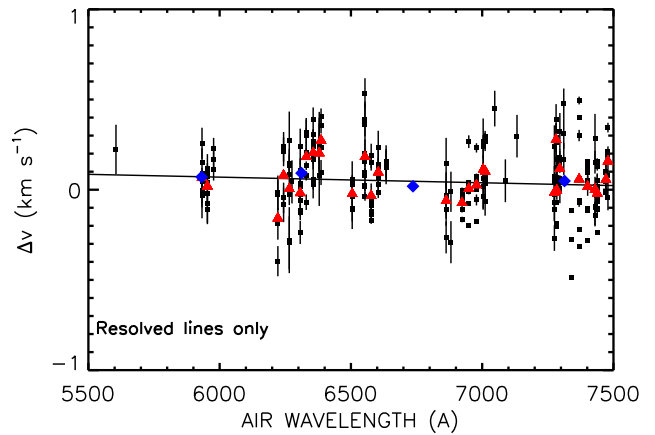}
  \caption{Black squares:  Offsets of the stronger sky lines in the individual
spectra from the wavelength position expected from the
sky-corrected ThAr calibration averaged over calibrations taken before and
after observations. Only lines with $\Lambda$-splitting more than the instrumental resolution ($R = 72,000$) are included.  Where more than four obervations
exist for a given skyline we also show the average with
the larger red triangle. The blue diamonds show the offsets averaged in 500\AA\ bins.  The black curve shows a linear fit to all of the data points. \label{fig:sky_all_ave}
}
\end{figure}

\begin{figure}[h]
\figurenum{7}
  \includegraphics[width=3.5in,angle=90,scale=0.7]{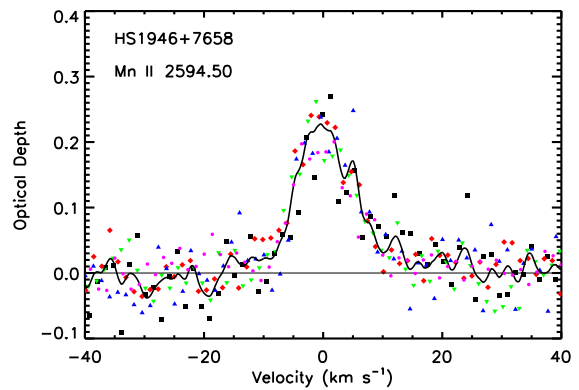}
  \caption{Comparison of the individual exposures  of
the \ion{Mn}{2} 2594.4976\AA\ line with the optimally
weighted average of all the spectra corresponding to a 10 hour exposure, 
which is shown by the black solid line. We have plotted
the optical depth versus the vacuum heliocenteric velocity
for a redshift of 1.7382. The individual exposures are shown without
smoothing or interpolation and each symbol shows an individual
pixel. The purple circles show the data from the September
2004 configuration (4 hours), the black squares the data
from July 2005 in the left configuration (80 minutes), the red
diamonds the left configuration from August 2006 (2 hours),
the blue upward pointing triangles the right configuration
from July 2005 (80 minutes) and the green downward pointing triangles
the right configuration from August 2006 (80 minutes).
\label{fig:vel_comp}
}
\end{figure}

We selected sky lines from this list with intensity $> 50$\ counts whose $\Lambda$-split components are resolved at our instrumental resolution and have roughly equal intensities.  We rejected lines whose FWHM from the Gaussian fit appeared to be too narrow, indicating that the line might be spurious.  This selection provided 20 -- 30 sky lines in each spectrum.

Figure~\ref{fig:sky_all_ave} shows the measured offsets in ${\rm km\ s}^{-1}$\ for these sky lines in all the spectra as a function of wavelength.  The offset is measured from the wavelength position expected from the final wavelength solution, as described above.  The black squares with error bars show the individual offsets, the red triangles the average offset in cases with four or more observations of a given sky line, and the blue diamonds show the offsets averaged in 500~\AA\  bins.  The solid line is a linear fit to all the points.  Over the range 5500 -- 7500\AA, where there are large numbers of sky lines, we find that there is a significant gradient, with $d(\Delta v)/d\lambda = (-2.91 \pm 0.66) \times 10^{-5}\ {\rm km\ s}^{-1}{\rm \AA}^{-1}$.  At $z = 1.5$, using \ion{Fe}{2} 2344\AA\ and \ion{Mg}{2} 2800\AA, this would result in an offset in the $\Delta\alpha /\alpha$\ measurement of $-1.7 \times 10^{-6}$, a significant correction to measured offsets.  Therefore a measurement of $\Delta\alpha /\alpha$\ with HIRES using \ion{Fe}{2} and \ion{Mg}{2} lines and ThAr calibration will show a negative bias.  \citet{griest10} have also found large systematic errors in the HIRES wavelength calibration between 5000~\AA\  and 6000~\AA, using the Iodine cell to track wavelength shifts, and  \citet{whitmore10}, using similar methods,  report similar though smaller systematics for the VLT/UVES spectrograph.  By comparing solar and asteroid spectra, \citet{rahmani13} found wavelength-dependent offsets for UVES, though with the opposite slope to our result for HIRES.

In what follows we used the average ThAr line measurement for the wavelength solution.  In Section~\ref{sec_mg2} we then apply the correction to the wavelength solution based on the sky line fits to correct the measured results so that we can see the comparative size of the effect.   
The wavelengths were then converted to vacuum 
heliocentric which is used, except where specified otherwise,
in the subsequent text.

\subsection{Testing the measurement errors}
\label{errortest}

Error estimation for both Voigt fitting and cross-correlation is complex
 and there is considerable room for systematic effects. It is 
our assessment that many of the previous analyses have assigned 
surprisingly, and possibly unrealistically, small errors to their results.
We estimated errors on our fitting in three ways. In the first
we measured independent observations of the lines to compare
the dispersion with the formal statistical error. 
In the second, we measured the scatter in an individual order using measurements of doublets in the spectra. 
In the third
we placed the fitted model for a given line 
at a large number of neighboring random positions in the spectrum 
and then profile-fitted these simulated lines.

%
%

\begin{figure}[h]
\figurenum{8}
\includegraphics[width=3.7in,angle=0,scale=1.0]{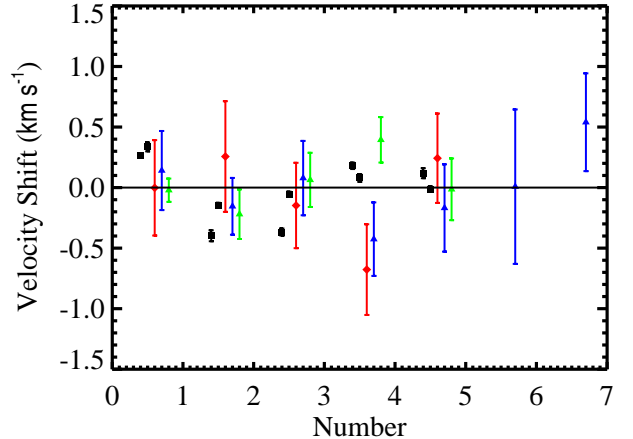}
\caption{Statistical and systematic errors from individual measurements of the $z = 1.7382$\ system in HS 1946+7658.  The velocity shift from lines of \ion{Fe}{2} 2600 (red diamonds), \ion{Fe}{2} 2382 (blue triangles), \ion{Mg}{1} (green triangles), and \ion{Mg}{2} (black squares) are shown for each available measurement.  The error bars are $1\sigma$\ statistical errors. 
\label{fig:systematic_errors}
}
\end{figure}

 The formal statistical error will not  include systematics, 
such as relative errors in the wavelength determination, uncertainties 
in the absorption line template, and so on. We therefore refitted the 
lines, separating the data by night of observation, 
and also separating systems that appeared 
in more than one order of the echelle spectrum. 
A comparison of the individual observations for a \ion{Mn}{2} absorption
line in HS1946+7658 with the final combined profile is shown in
Figure~\ref{fig:vel_comp}. Up to eight  independent measurements of the offsets 
were obtained in this 
way for some of the absorption lines, and these can be used to make 
an independent error estimate which should include nearly all 
of the these random systematic effects associated with 
the wavelength calibration and profile fitting (though 
not systematic effects that are present in all the observations such 
as the linear offset from the ThAr line calibration noted in
 the previous section).

In Figure~\ref{fig:systematic_errors} we show an example of the measured velocity
offsets in a single system relative to the value measured
in the combined spectrum. We show the \ion{Mg}{2} doublet (black
squares) and the \ion{Fe}{2} 2344, 2382 and 2600 lines as colored
symbols. For the weaker \ion{Fe}{2} lines, with their higher statistical
errors, the measured dispersion is fully consistent with the
statistical error. However for the strong \ion{Mg}{2} lines the statistical
error is small and is dominated by the systematic error which
is approximately $0.1~{\rm km\ s}^{-1}$  in each individual measurement. This
has little effect on any of the measurements discussed in the
next sections since the differential errors are always dominated
by weaker lines.

\begin{figure}[h]
\figurenum{9}
\includegraphics[width=3.7in,angle=0,scale=1.0]{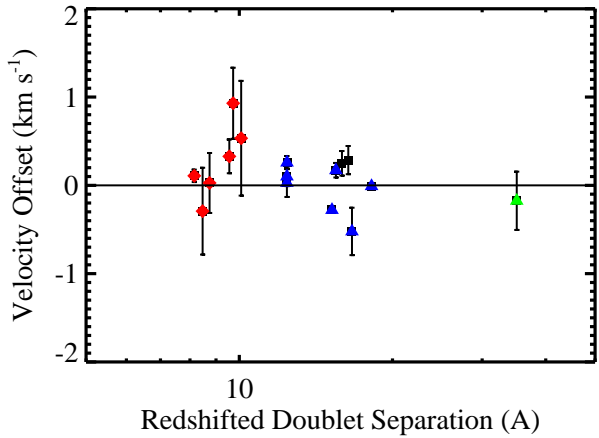}
\caption{Velocity offset in the doublet separation of \ion{C}{4} (red diamonds), \ion{Mg}{2} (blue triangles), \ion{Cr}{2} (black squares) and \ion{Si}{4} (green triangle) doublets as a function of redshifted doublet separation.  
\label{fig:sawtooth}
}
\end{figure}

Griest et al.'s Iodine cell observations suggest that much of the velocity uncertainty of the Th Ar calibration arises in individual orders (the sawtooth pattern seen in their Figure 4).  This is of considerable interest for methods such as the comparison of Cr and Zn lines discussed in section~\ref{sec_cr_zn_mn}.  We therefore tested this by measuring the doublet separations of \ion{C}{4}, \ion{Si}{4}, \ion{Mg}{2} and \ion{Cr}{2} lines in our data.  We used only lines where both members of the doublet were in a single order, where lines appeared to be a single symmetric component, and where both members of the doublet were strong but unsaturated.  We show the measured offsets as a function of the wavelength separation of the doublet at the observed redshift in Figure~\ref{fig:sawtooth}.  The highest 
S/N measurements again show a larger scatter than would be expected based on statistical errors, with a typical maximum error of $\sim 0.3\ {\rm km\ s}^{-1}$, consistent with the value expected from the Griest et al. result.  This error is expected to be random and should average out with a large enough number of measurements. The systematic error contributes a dispersion of about $0.17~{\rm km\ s}^{-1}$ in addition to the statistical error, which is large enough to substantially increase the error estimates.

%
%

%
%

\begin{figure}[h]
\figurenum{10}
\includegraphics[width=3.9in,angle=0,scale=0.9]{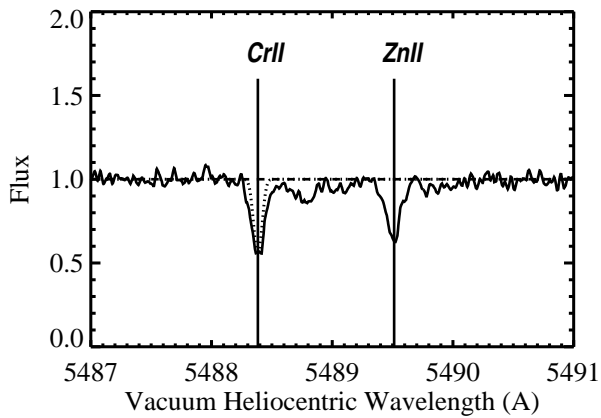}
\caption{An expanded region around the
\ion{Cr}{2} 2062.236 and the \ion{Zn}{2} 2062.660 lines showing
the structure of the absorption system at $z = 1.6614$\  in HE 1104$-$1805A. The normalized continuum is
shown by the dashed line and the instrument profile by the dotted
line. The feature of primary interest is the sharp line at the marked
wavelength
measured from the \ion{Cr}{2} line. This lies
at the blue edge of the absorption structure and is just resolved at
this resolution.  There is weaker broad absorption to the red of this
feature. The \ion{Zn}{2} line is cleanly separated from the
neighboring \ion{Cr}{2} absorption and also from the very weak \ion{Mg}{1}
2026 absorption which lies redward of it. We have also not identified
any contamination of our lines by absorption features arising from
other systems, though in any individual line measurement this remains as a
concern.
\label{fig:lines2}
}
\end{figure}

The simulation technique, which was also used by Chand et al. (2004), 
has the advantage of automatically including the effects of continuum fitting
 and line contamination in the spectrum as well 
as the effects of non-Gaussian noise, and is therefore 
considerably superior in these respects to estimates that rely on
 only the local statistical noise to estimate the error. We expect
{\it a priori} that this method will produce a larger estimate of the noise because
of these effects. It also allows us to quantify the fraction of times 
a line will suffer a major error because of contamination
 by another absorption line feature or other structures in the the spectrum.
We find that this type of analysis gives an average increase in the noise 
by a factor of 1.17 over the formal statistical errors, with a
 range from 0.9 -- 1.3. 
 This is slightly smaller than the factor of 1.5 seen by \citet{agafonova11}.  
 Most contamination by other lines
is easily recognized and would be excluded in the actual measurements,
but significant velocity offsets ($> 4\sigma$) which would not be easily recognized
 are seen in about 1\% of the simulated lines.

\section{\ion{Cr}{2}, \ion{Zn}{2} and \ion{Mn}{2} measurements}
\label{sec_cr_zn_mn}

As we have discussed above, measurements based on the \ion{Cr}{2}, \ion{Zn}{2} and \ion{Mn}{2} lines can avoid some of the wavelength calibration problems though they are still subject to the in-order errors.  However, as we discuss here, variability in relative abundances in the substructure of the line can cause serious problems in analyzing even apparently weak and simple systems.  
A portion of the spectrum of HE1104-1805A
covering the \ion{Zn}{2} and \ion{Cr}{2} lines of interest in an absorption system at $z = 1.6614$\  is shown in
Figure~\ref{fig:lines} and an expanded region around the \ion{Cr}{2} 2062.236 and \ion{Zn}{2} 2062.660 lines is shown in Figure~\ref{fig:lines2}.  The expanded spectrum demonstrates that
the wavelength structure of this absorption is indeed very simple in
the weaker singly-ionized lines and consists of a single sharp feature
which is only marginally resolved, together with very weak broader
absorption to the red, which extends only to about $50~{\rm km\
s}^{-1}$ and is nearly fully separated from the sharp feature. It is the sharp singly-ionized line which we focus on here.
The velocity offset between the \ion{Cr}{2} and \ion{Zn}{2} lines can already be seen by
visual inspection of Figure~\ref{fig:lines2}.

We first fitted Voigt profiles to all of the useful singly-ionized
lines that were not saturated. We used our own customized IDL routines, based on the MPFIT programs of \citet{markwardt09}.  
(We note that a
cross-correlation analysis gives basically identical results).  A
two-component fit based on the strong \ion{Si}{2} 1808 line was
used. The principal component, containing the bulk of the material, is
the narrow feature with a $b$-value of $4.1~{\rm km\ s}^{-1}$; a single broad component was used to model the weaker red wing. We
fitted each of the lines by holding the $b$-values fixed but allowing
the column densities of all the components, and the overall velocity, to vary. The redshifts of
the sharp component were measured and the corresponding velocity
offsets  determined. The offsets are not
sensitive at any significant level to the choice of cloud model, and the results are almost unchanged if we allow the column density, velocity and $b$-value to vary independently in the two components.  We note that the velocities are consistent between independent measurements of a given ion, suggesting that line contamination and line saturation are not problems.

%
%

\begin{figure}[h]
\figurenum{11a}
\includegraphics[width=3.7in,angle=0,scale=1.0]{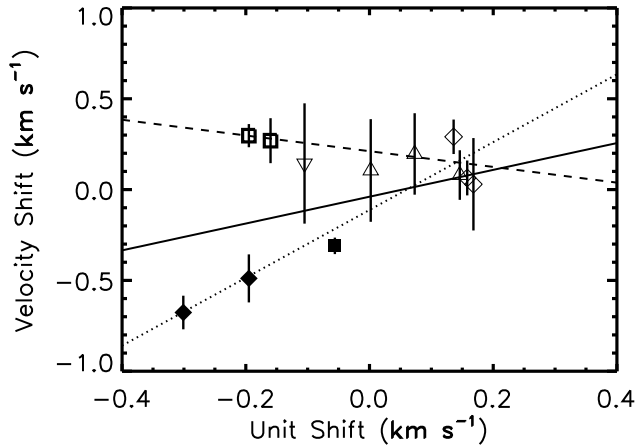}
\caption{The measured velocity offsets from Voigt profile
analysis in the various absorption lines in the $z=1.6614$ 
system of
HE1104-1805A are plotted against the offset which would be generated
by a change in $\alpha$ of 10$^{-5}$.  The symbols show the measured
values with $1~\sigma$ errors (statistical only).  The filled diamonds are
the \ion{Zn}{2} 2026 and 2062 lines, the filled square the \ion{Si}{2} 1808 line, the open 
upward triangles the \ion{Ni}{2} 1709, 1741 and 1751 lines, the open diamonds
the \ion{Cr}{2} 2056, 2062 and 2066 lines, the downward triangle the \ion{Fe}{2}
1611 line and the open squares the \ion{Mn}{2} 2576 and 2594 lines. The solid
line shows the optimal linear fit to all of the lines, which corresponds to a 
$\Delta\alpha /\alpha$ of $(0.74\pm0.20) \times 10^{-5}$ 
(statistical error only).  The dotted line shows the fit obtained by including only the \ion{Zn}{2} and \ion{Cr}{2} lines, giving $\Delta\alpha /\alpha$ of $(1.87\pm0.24) \times 10^{-5}$.  
The dashed line shows the null result of $\Delta\alpha /\alpha = (-0.43\pm0.24) \times 10^{-5}$, 
obtained by including only the \ion{Mn}{2}, \ion{Ni}{2} and \ion{Cr}{2} lines.  Wavelengths are from the compilation of \citet{murphy14}.
 The unit shift is the
displacement in velocity which would be produced by a
$\Delta\alpha/\alpha =10^{-5}$ from \citet{dzuba02}  
and \citet{berengut04}. 
\label{fig:velshift}
} 
\end{figure}

We show the relation between the offset expected from a change in
$\alpha$ and the measured velocity offset for the $z=1.6614$ system
in HE1104-1805A in  Figure~\ref{fig:velshift}. A
linear regression analysis gives  $(0.74\pm0.20) \times 10^{-5}$ 
(statistical error only) for the required $\Delta\alpha /\alpha$. The best fit is shown by
the solid line in the figure.  However, the fit has a very poor
$\chi^2$ because of the offset between the \ion{Mn}{2} and \ion{Zn}{2}
lines.  We have also measured the offset which would be obtained from
restricted subsets of the lines. If we include only the \ion{Zn}{2} and \ion{Cr}{2} lines we obtain $(1.87\pm0.24) \times 10^{-5}$, 
whereas if we use the \ion{Mn}{2}, \ion{Ni}{2} and \ion{Cr}{2} lines this reduces to a null result of $(-0.43\pm0.24) \times 10^{-5}$. 
Including the systematic errors discussed in \citet{griest10} and Section~\ref{errortest} would significantly  increase the errors from $0.24 \times 10^{-5}$\ to $0.46 \times 10^{-5}$ but still leave a highly significant difference between the fits based on the \ion{Zn}{2} and \ion{Mn}{2} lines. 

This result shows that there can be serious systematic problems in
determining $\Delta\alpha$ using measurements of just \ion{Cr}{2} and
\ion{Zn}{2}.  It strongly suggests that, even in a very simple narrow
system like the present one, abundance variation in substructure in
the sharp feature (the most likely systematic) can produce an apparent
offset between the species.  Zn and Cr are well known to have very
different abundance patterns, an effect
which has traditionally been ascribed to dust depletion but may also
reflect variations in nucleosynthesis \citep{pettini01,prochaska03}.  
If a positive subcomponent of
the narrow absorption line had a high Cr/Zn ratio relative to a
negative subcomponent this could pull the relative offsets of
\ion{Zn}{2} and \ion{Cr}{2} to negative values.  This is consistent
with the observed velocity pattern: the weakly refractory \ion{Zn}{2}
and \ion{Si}{2} have the most negative velocities, shown by the filled symbols in Figure~\ref{fig:velshift}, 
while the refractory
elements (\ion{Ni}{2}, \ion{Mn}{2}, \ion{Cr}{2} and \ion{Fe}{2}) have
more consistent positive velocities.   It appears that using \ion{Mn}{2} with \ion{Cr}{2} is more secure than using \ion{Zn}{2} with \ion{Cr}{2} since
both \ion{Mn}{2} and \ion{Cr}{2} are refractory elements.

Changes in the isotopic mix are unlikely to produce significant changes in the results.     
The
isotopic effect may be largest in the \ion{Zn}{2} lines but even here the
maximum shift which could be obtained by changing the isotope mix 
from the local uniform mix of isotopes to being predominantly $^{64}$Zn  is only about $0.1~{\rm km\ s}^{-1}$, which is small compared with the
measured shifts.   As long as $^{52}$Cr remains the dominant isotope we expect the \ion{Cr}{2} lines to have relatively small isotopic shifts \citep{murphy14}.

\begin{figure}[h]
\figurenum{11b}
\includegraphics[width=3.7in,angle=00,scale=1.0]{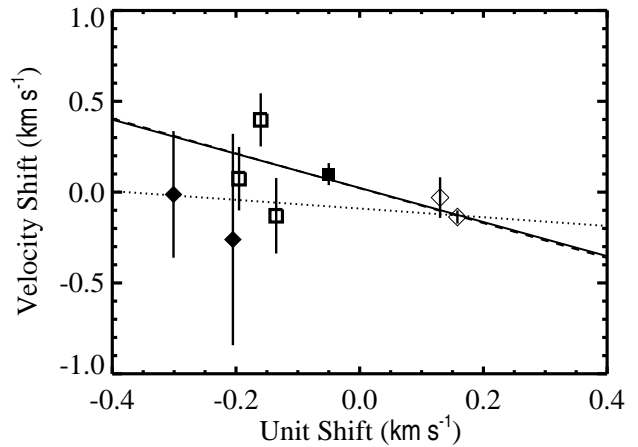}
\caption{As in ({\it a\/})
for the $z=1.7382$ system in HS1946+7658.  Here the fit to all the measured lines gives
$\Delta\alpha /\alpha = (-0.94\pm0.24) \times 10^{-5}$ 
(solid line), 
while fitting only \ion{Zn}{2} and \ion{Cr}{2} gives 
$\Delta\alpha /\alpha = (-0.24\pm0.69) \times 10^{-5}$ 
(dotted line), 
and fitting only \ion{Mn}{2} and \ion{Cr}{2} gives 
$\Delta\alpha /\alpha = (-0.96\pm0.32) \times 10^{-5}$ 
(dashed line).
\label{fig:velshiftb}
} 
\end{figure}

The $z=1.7382$ system has a very similar structure and we adopted the same fitting procedure.  The measured offsets 
shown in 
Figure~\ref{fig:velshiftb} yield a consistent solution for both
\ion{Zn}{2} and \ion{Mn}{2} relative to \ion{Cr}{2}, albeit with
larger errors for this weaker system.  The regression analysis gives
$\Delta\alpha/\alpha = (-0.94\pm0.24) \times 10^{-5}$ 
with an acceptable $\chi^2$, and consistent results between \ion{Zn}{2} and \ion{Cr}{2} and \ion{Mn}{2} and \ion{Cr}{2}.  Again, including the systematic error from Section~\ref{errortest} raises the error from $0.24 \times 10^{-5}$ to $0.59 \times 10^{-5}$.  
Adopting the values based on \ion{Mn}{2}, \ion{Ni}{2} and \ion{Cr}{2} ions and correcting for the systematic errors, we consider that the most reliable measurements are $\Delta\alpha/\alpha = (-0.47 \pm 0.53) \times 10^{-5}$\ 
for the $z = 1.6614$\ system in HE1104$-$1805A and $\Delta\alpha/\alpha = (-0.79 \pm 0.62) \times 10^{-5}$\ 
for the $z = 1.7382$\ system in HS1946+7658.  However, systematic errors remain a concern even when using the more consistent ions and in particular abundance variations may cause velocity shifts.  Large numbers of systems would be required to average out the latter effect.

%
%

%
%

\begin{figure}[h]
\figurenum{12a}
\includegraphics[width=3.5in,angle=0,scale=0.83]{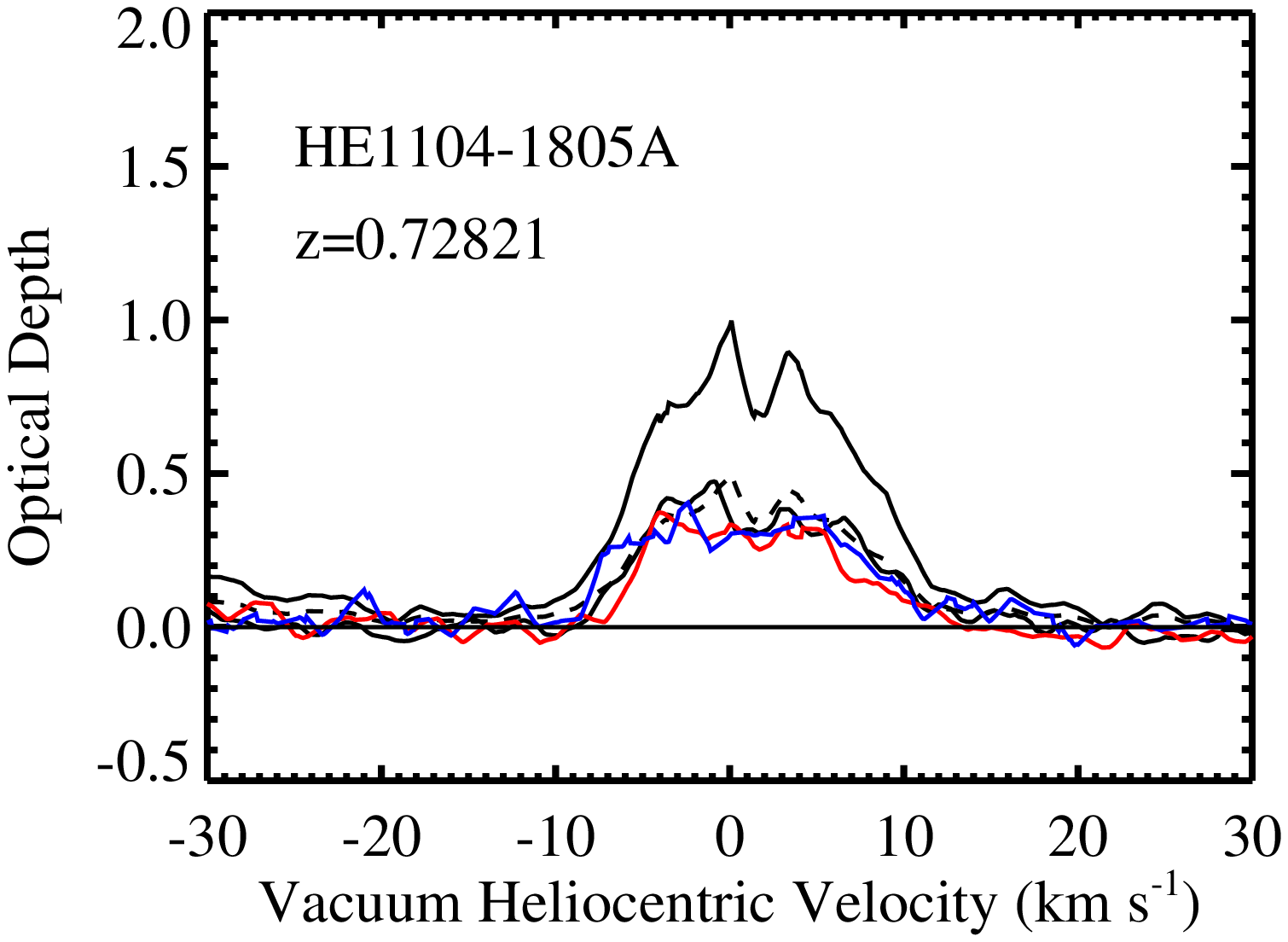}
\includegraphics[width=3.5in,angle=0,scale=0.83]{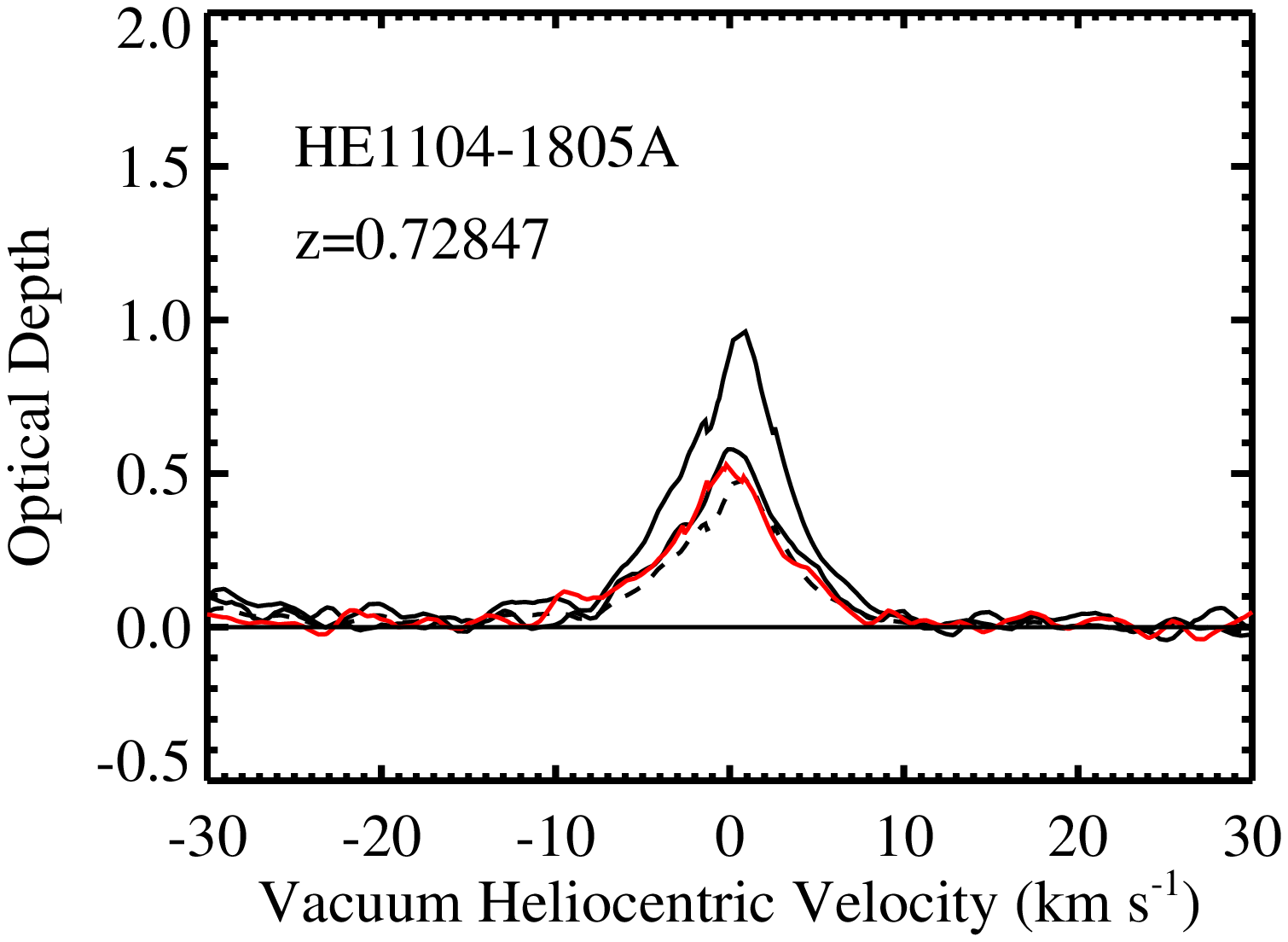}
\includegraphics[width=3.5in,angle=0,scale=0.83]{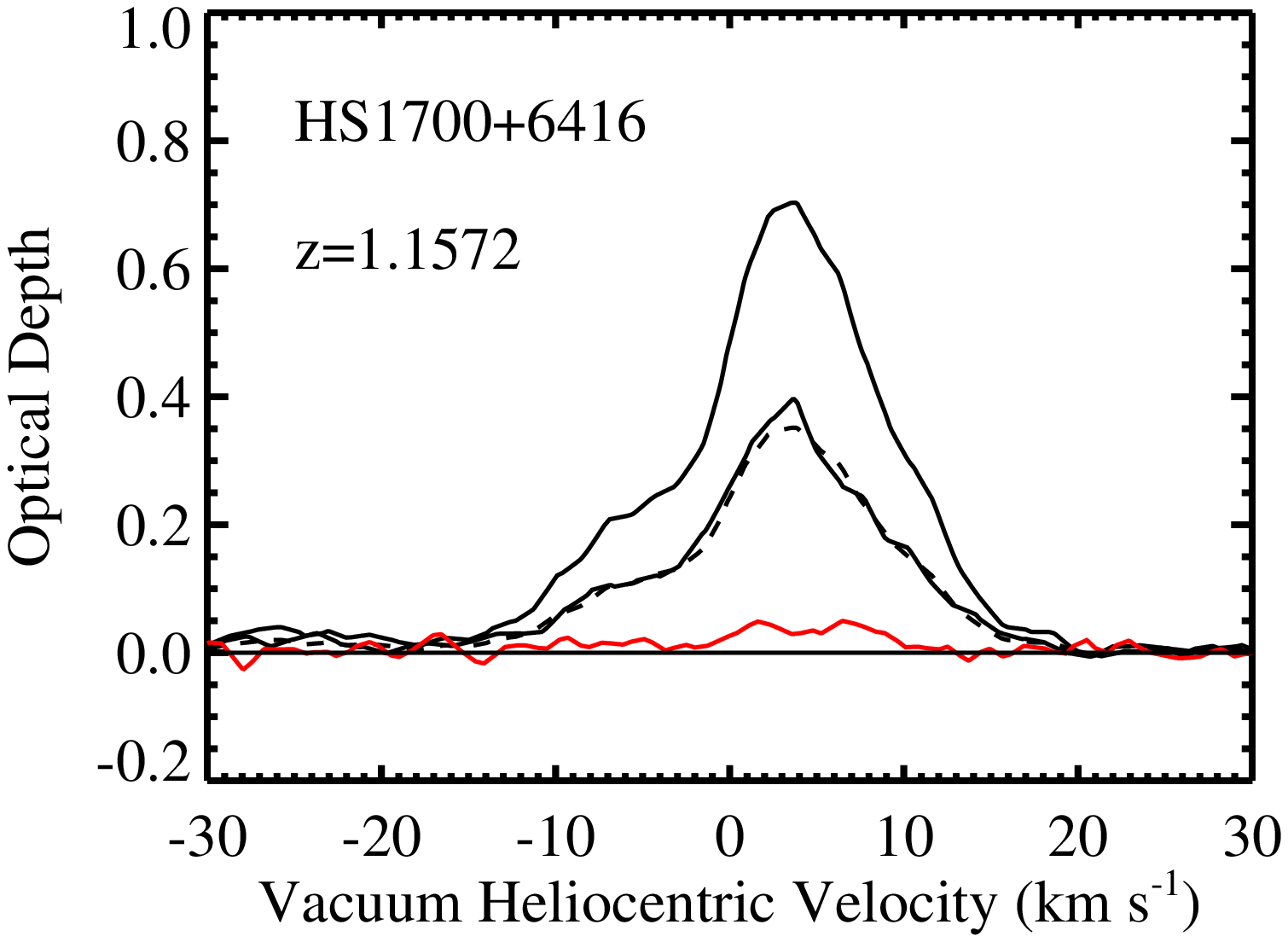}
\includegraphics[width=3.5in,angle=0,scale=0.83]{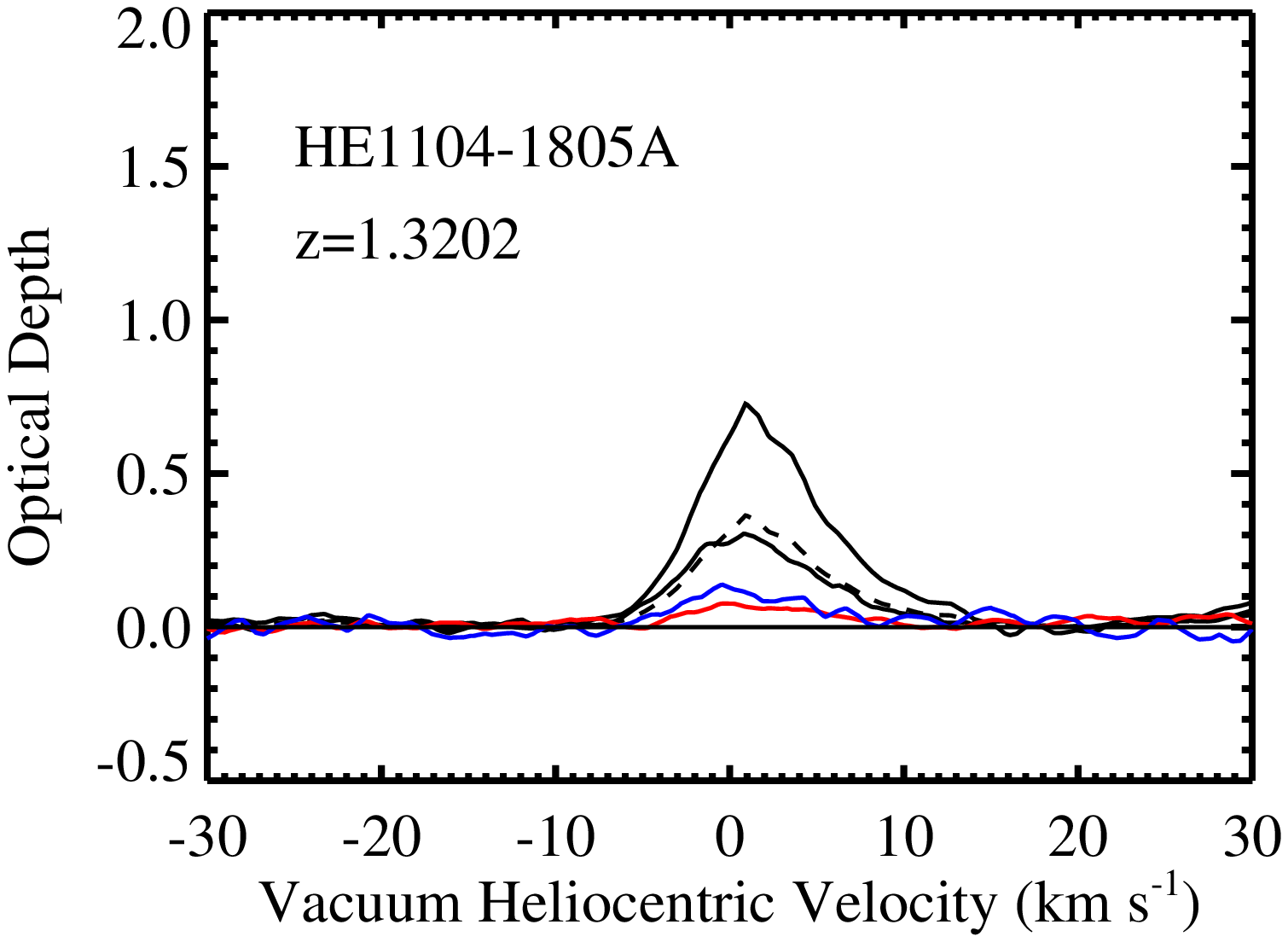}
\caption{Optical depth of \ion{Fe}{2} 2600 (red), \ion{Fe}{2} 2382 (blue) and the \ion{Mg}{2} doublet (black) for the four \ion{Fe}{2} - \ion{Mg}{2} systems that show no evidence of saturation in the \ion{Mg}{2} doublet.  The black dashed line shows the \ion{Mg}{2} 2796 optical depth divided by a factor of two.  (Note that the vertical scale is enhanced in one of the panels for clarity). 
\label{fig:fe_mg2_lines}
}
\end{figure}

%
%

\begin{figure}[h]
\figurenum{12b}
\includegraphics[width=3.5in,angle=0,scale=0.83]{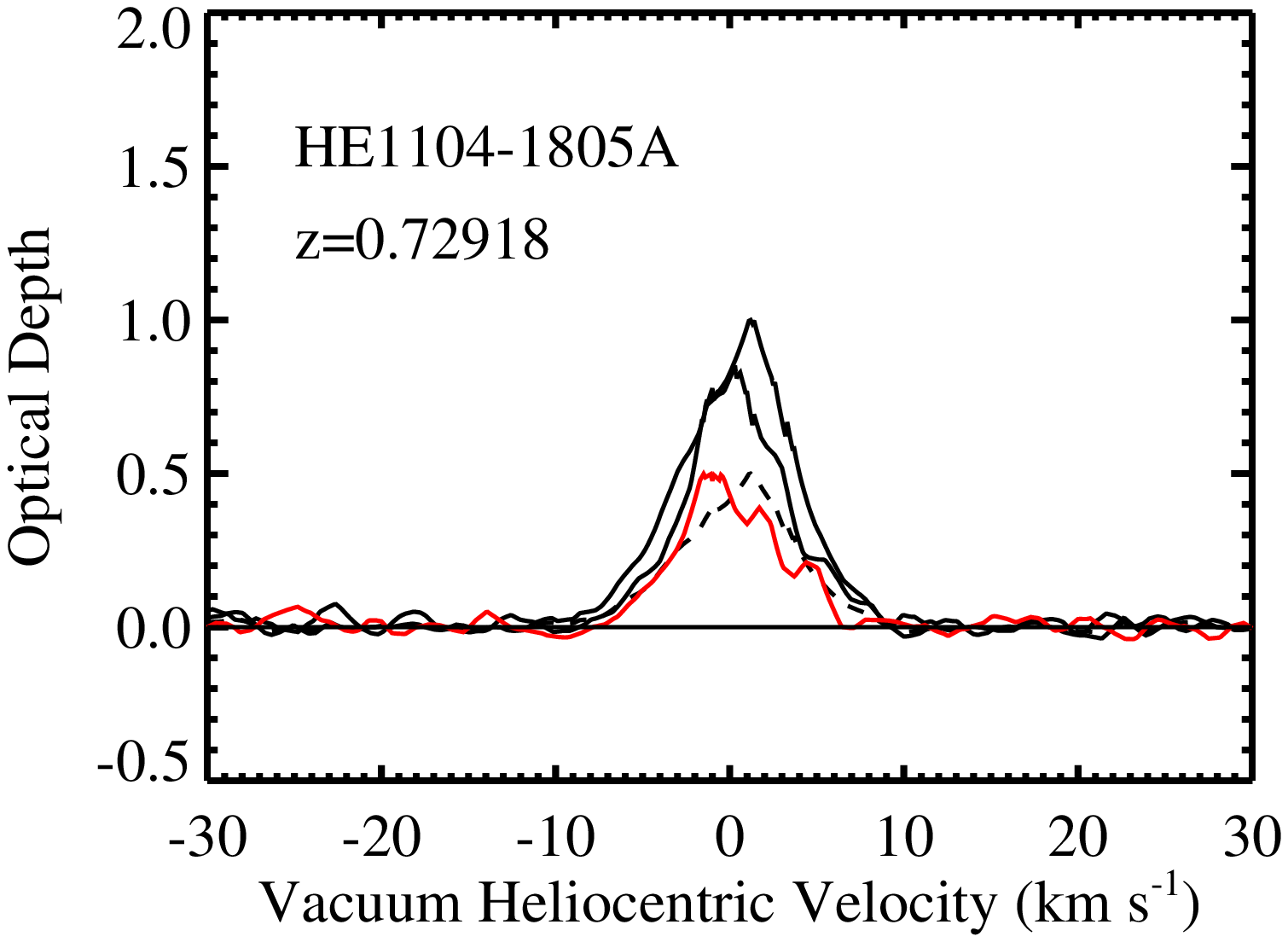}
\includegraphics[width=3.5in,angle=0,scale=0.83]{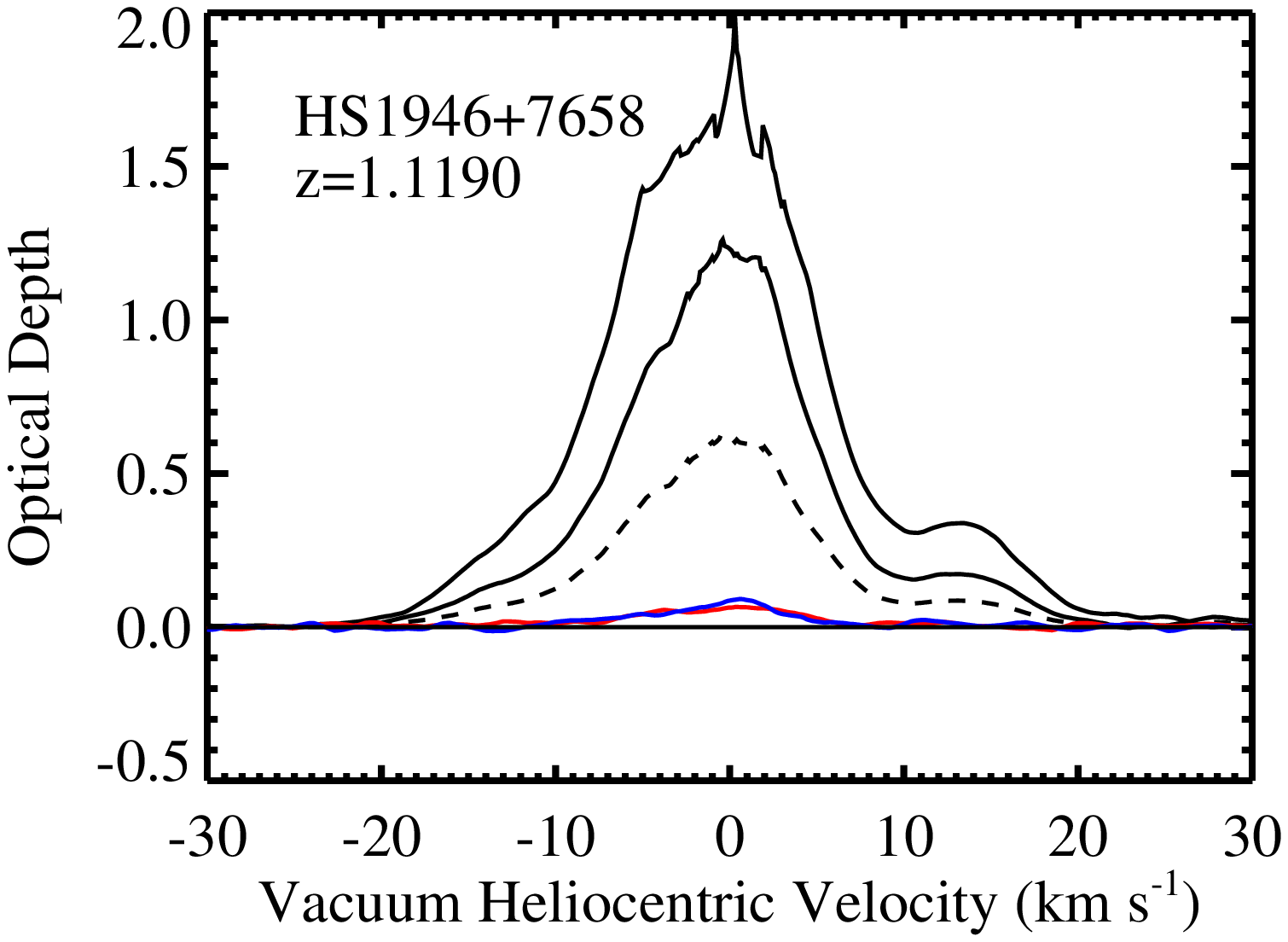}
\includegraphics[width=3.5in,angle=0,scale=0.83]{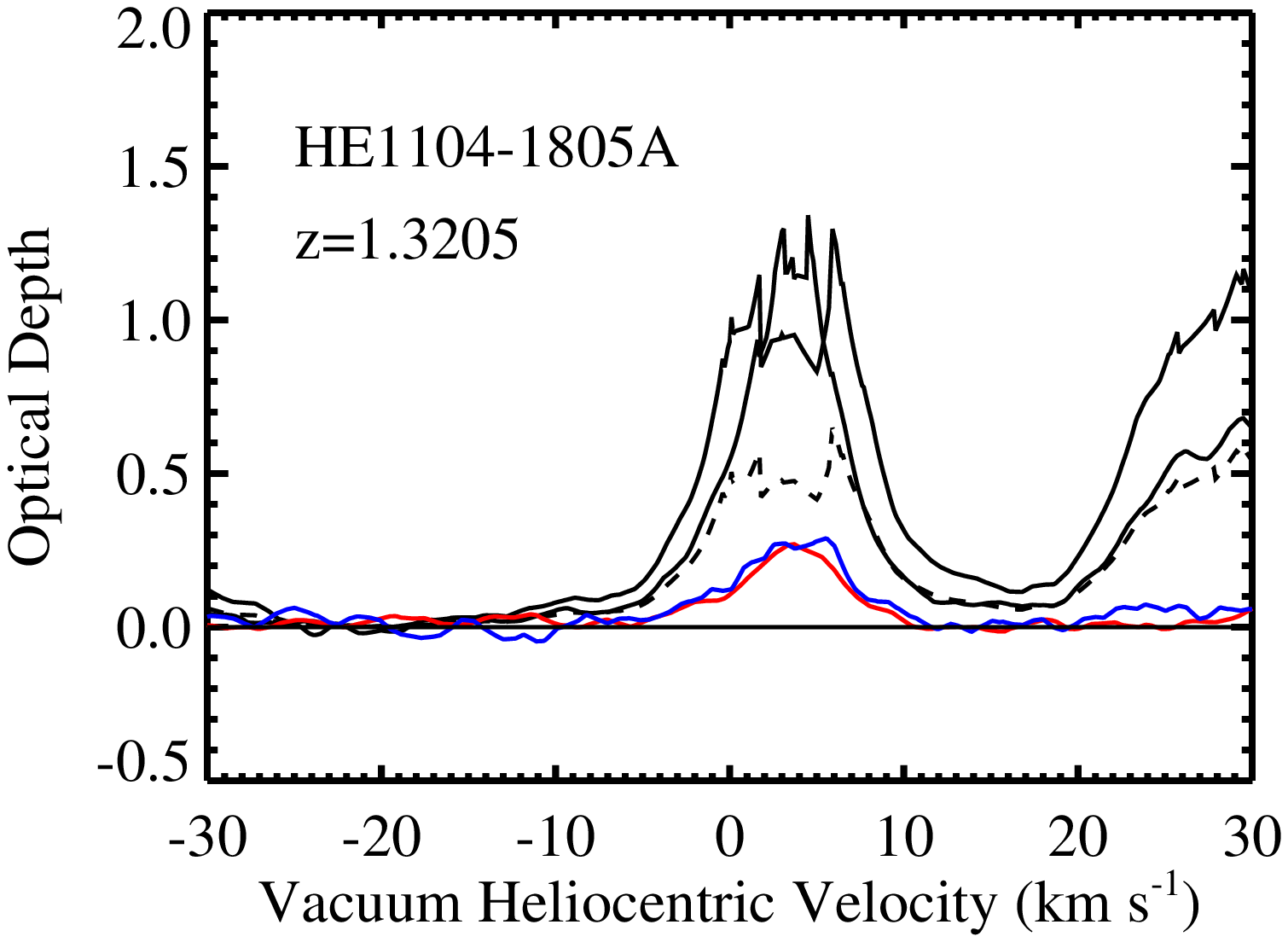}
\includegraphics[width=3.5in,angle=0,scale=0.83]{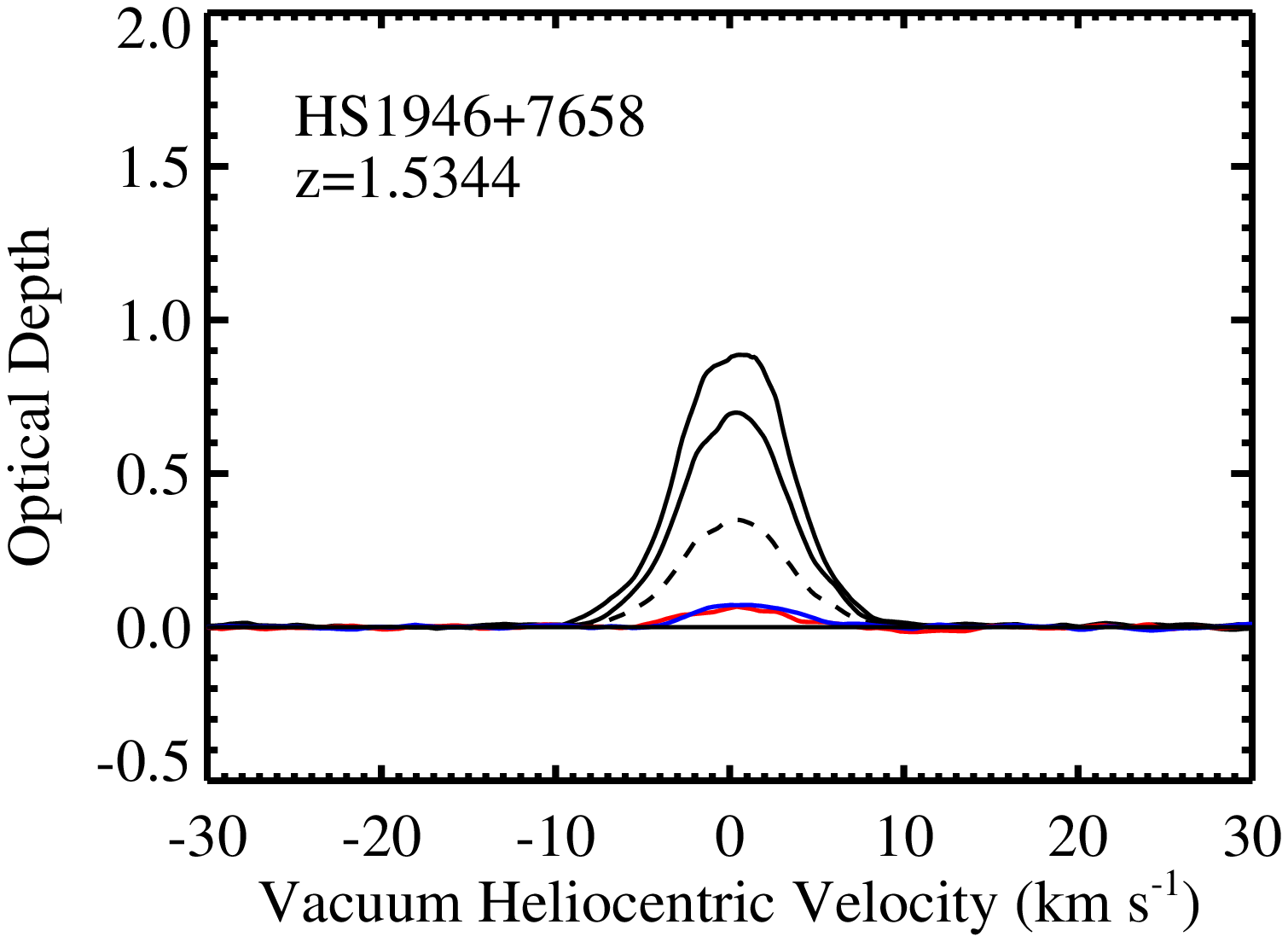}
\caption{As in (a) for the four \ion{Fe}{2} - \ion{Mg}{2} systems that show some evidence of saturation in the \ion{Mg}{2} doublet.\label{fig:fe_mg2_lines}
}
\end{figure}

\section{\ion{Mg}{2} measurements}
\label{sec_mg2}

\begin{figure}[h]
\figurenum{13}
\includegraphics[width=3.7in,angle=0,scale=1.0]{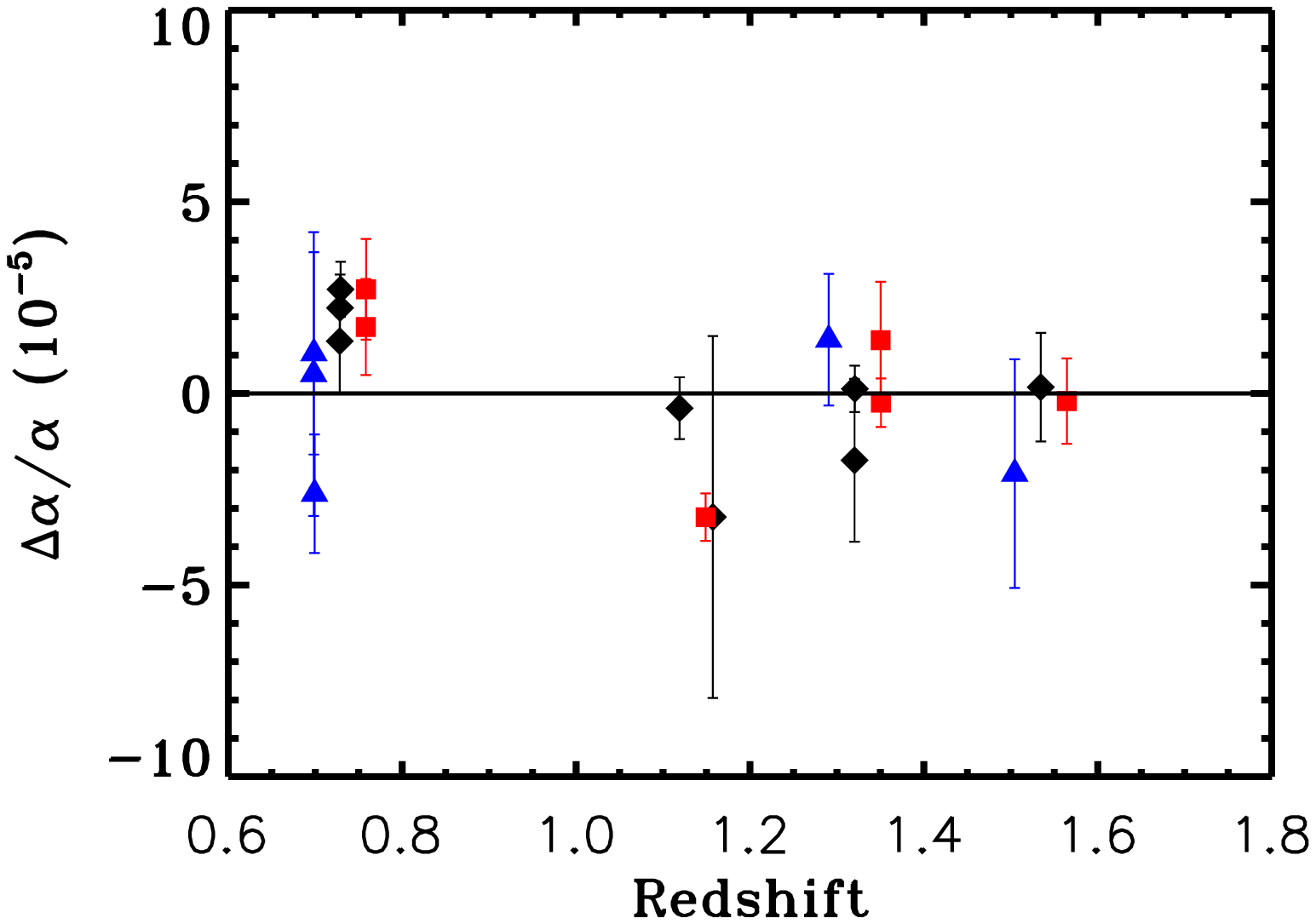}
\caption{Measured  $\Delta\alpha/\alpha$ based on the offsets of the Fe lines from the
\ion{Mg}{2} 2803 line are shown versus redshift. 
\ion{Fe}{2} 2600 measurements are shown with black diamonds, \ion{Fe}{2} 2344 with blue triangles, and \ion{Fe}{2} 2382 with red squares.  
The bars show the $1~\sigma$ statistical errors. The \ion{Fe}{2} 2344 and \ion{Fe}{2} 2382 lines are slightly displaced in redshift for clarity.  The weighted mean value is $\Delta\alpha/\alpha = (0.10 \pm 0.24) \times 10^{-5}$.
\label{fig:z_show}
} 
\end{figure}

We next analysed the eight narrow \ion{Mg}{2} absorption systems seen in
the three quasars. These are drawn from five independent complexes,
two of which contain two or three useful isolated narrow features.  The \ion{Fe}{2} and \ion{Mg}{2} lines in each of the systems are shown in Figure~12.  
In all
cases the lines are separated from neighboring absorption features
and at least some of
the \ion{Fe}{2} lines are strong enough for accurate measurements.  
The \ion{Mg}{2} doublet ratio can be used to estimate the degree of saturation, and there are four systems in which the \ion{Mg}{2} lines appear unsaturated and four in which at least the \ion{Mg}{2} lines show saturation effects.  
We profile-fitted each of the
lines using a model based on the \ion{Mg}{2} 2803 line and varying only the individual column densities of the components and the overall velocity of the system. The results are shown in Figure~\ref{fig:z_show}.


Figure~\ref{fig:z_show} shows that the spread of these systems,
which range in redshift from $z=0.7$\ to $z=1.5$, is consistent with
the assigned errors. Including only statistical errors, the error-weighted $\Delta\alpha/\alpha$\ is $(0.10 \pm 0.24) \times 10^{-5}$. 
When we correct for the wavelength bias discussed in Section~\ref{skycal} and for the systematic errors discussed in Section~\ref{errortest} this becomes $\Delta\alpha/\alpha = (0.43 \pm 0.34) 
 \times 10^{-5}$. 
The small offsets for \ion{Mg}{2} relative to \ion{Fe}{2} also show that change in  the isotopic mix of \ion{Mg}{2} from the local value is not large in these systems and is not weighted toward the heavier isotopes.  However, the effects of the isotopic mix in some of the quasar absorption lines can be large \citep{agafonova11} and could contribute to the spread seen here.

\section{Summary}
\label{secsummary}
 
MM measurements of the evolution of the fine structure constant are generally based on
either comparison of \ion{Mg}{2} and \ion{Fe}{2} lines or (at higher redshifts) on measurements
of \ion{Cr}{2}, \ion{Zn}{2}, \ion{Ni}{2}  and \ion{Mn}{2} lines. Both have significant problems which have to be
dealt with in order to obtain an accurate limit. 

In the case of the widely separated \ion{Mg}{2} and \ion{Fe}{2} lines, extreme care must
be taken in the wavelength calibration. As has previously been discussed in
\citet{griest10},  
we see drift in the HIRES wavelength calibration with time
together with a significant systematic error in the ThAr calibration of each order of the spectrum 
and, using the sky lines as a calibrator, a significant linear deviation
between the sky line based calibration and the ThAr line calibration. The
latter effect will produce a negative deviation in the measurement of 
$\Delta\alpha/\alpha$   for HIRES observations. The exact shift depends on the
redshift and line pair used but is around $1.7 \times10^{-6}$ for typical observations
and would reduce the significance of the \citet{murphy03a}
evolution
from $5\sigma$ to about $3.5\sigma$, assuming that the sky offset on HIRES remains at a fixed level over the course of time and has not changed between the observations of \citet{murphy03a} and those presented here.
Similar effects have been noted for the UVES spectrograph on the VLT \citep{griest10, whitmore10,rahmani13, bonifacio14}.  In particular, \citet{rahmani13} found wavelength-dependent offsets of a similar magnitude to those discussed here for HIRES by comparing solar and asteroid spectra in the wavelength range 3300\AA -- 3900\AA.  However, these offsets vary with wavelength in the opposite sense to the slope we find for HIRES, and so will produce a positive deviation.  If a consistent negative slope is found in further tests of the HIRES wavelength accuracy, the opposite signs of the HIRES and UVES observations could provide a natural explanation of the apparent spatial variation in $\alpha$\ claimed by \citet{king12}.  
 
We have measured the variation using our present high S/N and high resolution
observations of eight systems in  \ion{Mg}{2} and \ion{Fe}{2}, obtaining a value of $\Delta\alpha/\alpha = (0.43 \pm 0.34) \times 10^{-5}$ 
when we correct for the systematic bias and allow
for systematic errors in the line fitting. This null result is consistent
with most previous measurements using a variety of techniques. 

Measurements based on the neighboring lines of \ion{Cr}{2}, \ion{Zn}{2} etc.  occur in higher column density systems where abundance variation
in unresolved subcomponents can cause measurement errors. In the present work we
have argued that \ion{Mn}{2}, \ion{Ni}{2} and \ion{Cr}{2} may give the most robust results
for $\Delta\alpha/\alpha$ 
and that \ion{Zn}{2} and \ion{Cr}{2} measurements should be avoided. Using the \ion{Mn}{2}
and \ion{Cr}{2} measurements  for two weak and narrow systems we found limits
of $(-0.47 \pm 0.53) \times 10^{-5}$ 
and $(-0.79 \pm 0.62) \times 10^{-5}$ 
for $\Delta\alpha/\alpha$ when we allow for systematic wavelength calibration errors.

Combining the \ion{Mg}{2} and \ion{Fe}{2} results  with the \ion{Cr}{2} and \ion{Mn}{2} results
gives us a final null result of $\Delta\alpha/\alpha = 
(0.01 \pm 0.26) \times 10^{-5}$.  
We conclude that spectrosopic measurements, made using
these methods, are not yet capable of unambiguously detecting
variation in $\alpha$.

\acknowledgements

We gratefully acknowledge support of this work through the National
Science Foundation.  We would like to thank Amy Barger for drawing our
attention to work on the spatial variation of the fine structure
constant and to acknowledge Maria Pereira's early contributions to our
cross correlation analysis of the quasar absorption lines.  We would also like to thank the anonymous referee and Michael Murphy, John Webb, and Sergei Levshakov for comments on the first draft of the paper.


\clearpage

\end{document}